\documentclass[aps,nofootinbib,superscriptaddress,twocolumn]{revtex4}
\usepackage{float}
\usepackage{graphicx}  
\usepackage{subfig}
\usepackage{amsmath}   
\usepackage{amssymb}   
\usepackage{bm} 
\usepackage{dcolumn}
\usepackage{color}
\usepackage{mathrsfs}
\usepackage{amsfonts}
\usepackage{varioref}
\usepackage{textcomp}
\RequirePackage[colorlinks,citecolor=blue,urlcolor=magenta,linkcolor=blue]{hyperref}
\allowdisplaybreaks

\labelformat{section}{Section #1} 
\labelformat{subsection}{Section #1} 
\labelformat{subsubsection}{Section #1}
\labelformat{subsubsubsection}{Section #1}
\labelformat{equation}{Eq.~(#1)} 
\labelformat{figure}{Fig.~#1} 
\labelformat{subfigure}{Fig.~\thefigure#1} 
\labelformat{table}{Tab.~#1} 
\begin{document}

\title{Asymptotically de-Sitter black holes have non-zero tidal Love numbers}

\author{Sreejith Nair}
\email{sreejithnair@iitgn.ac.in}
\affiliation{Indian Institute of Technology, Gandhinagar, Gujarat-382355, India}

\author{Sumanta Chakraborty}
\email{tpsc@iacs.res.in}
\affiliation{School of Physical Sciences, Indian Association for the Cultivation of Science, Kolkata-700032, India}

\author{Sudipta Sarkar}
\email{sudiptas@iitgn.ac.in}
\affiliation{Indian Institute of Technology, Gandhinagar, Gujarat-382355, India}

  
\begin{abstract}
Love numbers of compact objects quantify their tidal deformability against external perturbations. It is expected that Love numbers of asymptotically flat black holes (BHs) in General Relativity are identically zero. We show that quite contrary to common expectations, the tidal Love numbers of asymptotically de Sitter black holes are non-zero.

\end{abstract}
\maketitle
    \section{Introduction}
    Gravitational wave (GW) observations have opened up a new frontier for testing Einstein's general relativity and any possible modification to the same \cite{LIGOScientific:2016aoc,LIGOScientific:2016lio}. Gravitational wave observations can potentially test general relativity at the strong field regimes, where we expect to see possible signatures of new gravitational physics.\\

    One of the captivating implications of Einstein's general theory of relativity (GR) is the existence of configurations with an event horizon, representing a causal boundary such that any event within the confines of an event horizon cannot exert causal influence on events outside it. GW observations serve as a valuable tool for testing the existence of astrophysical black holes with event horizons \cite{LIGOScientific:2016lio,Agullo:2020hxe,Cardoso:2017cfl,Maggio:2021uge, Nair:2022xfm,Katagiri:2023yzm,Cardoso:2019nis,Chakraborty:2022zlq,Cardoso:2022fbq,Datta:2019epe,Saketh:2022xjb,Cardoso:2019rvt,Katagiri:2023umb,Chakraborty:2023zed,DeLuca:2022tkm}.\\ 

    An important observation with regards to the black hole solutions of general relativity (GR) is that they have zero tidal deformability \cite{Zhang:1986cpa,Poisson(Book):2014,Hinderer:2007mb,Damour:2009vw,Kol:2011vg,Pani:2015hfa,Landry:2015zfa,Binnington:2009bb,Chia:2021,Ivanov:2022hlo,Addressingissues:2023,Creci:2021rkz,Ivanov:2022qqt,Hui:2020xxx}. The tidal deformability being quantified through the linear response of the multipole moments of the compact object to an external tidal field. The real part of the constant quantifying the linear response is called the Love number \cite{Zhang:1986cpa,Poisson(Book):2014,Binnington:2009bb,Hinderer:2007mb}. Love numbers of compact objects as measured through the GWs emitted from binaries is regarded as a powerful tool to test for black holes and thus potential deviations from Einstein gravity \cite{Cardoso:2017cfl,DeLuca:2022tkm,Nair:2022xfm,Katagiri:2023yzm,Katagiri:2023umb,Chakraborty:2023zed}. The vanishing of static black hole Love numbers of the Kerr family of black holes in Einstein gravity are also of significant interest from a purely theoretical perspective. This is due to its association with the no-hair theorem and with certain symmetries of the spacetime, resulting in a Ladder structure which can be used to relate different modes of perturbations on the black hole background \cite{Hui:2021vcv,BenAchour:2022uqo,Berens:2022ebl,Sharma:2024hlz}.\\
    
    This work establishes that not all black holes in GR have zero tidal Love numbers. We demonstrate that asymptotically de Sitter (dS) black holes within Einstein's gravity with a mass $M$ and cosmological constant $\Lambda$ have a non-zero scalar Love number at $\mathcal{O}(\Lambda M^2)$. This has many significant consequences, as it demonstrates that even in GR, objects with horizons can have a non-zero Love number. This is besides the fact that current observations suggest our universe to be de Sitter. As a result, the asymptotically dS black holes of GR, known as Schwarzchild de Sitter (SdS) black holes, may be more observationally relevant than the asymptotically flat family of black holes.\\ 

    For computing the Love numbers of SdS black holes, we will use the worldline Effective Field Theory (EFT) \cite{Goldberger:2004jt,Kol:2011vg, Creci:2021rkz}, originally developed for asymptotically flat compact objects, adapted for comoving asymptotically de Sitter compact objects. At the level of the macroscopic worldline EFT \cite{Kol:2007rx,Chakrabarti:2013xza,Ivanov:2022qqt,Mandal:2023hqa,Bhattacharyya:2023kbh}, the characteristic length scales of the extended bodies are integrated out, allowing us to treat them as point particles, moving on the background spacetime. The finite-size effects are accounted for through the coupling of additional fields on the point particle world line. In the context of the dS worldline EFT, we will note that the necessary calculations are more transparent in the conformally flat coordinates of the Poincaré patch. As a result, we will construct the worldline EFT for extended bodies on a de Sitter background in the conformally flat coordinates of the Poincaré patch.\\
    
    We will use the scattering amplitudes of scalar fields within the dS worldline EFT framework as measured by a far away comoving observer on the de Sitter background to define Love numbers. We adopt this approach to define Love numbers, as the conventional approaches \cite{Poisson(Book):2014,Binnington:2009bb} cannot be used if the spacetime is not asymptotically flat. So, our formalism extends the definition based on scattering coefficients from worldline EFT in flat spacetime \cite{Creci:2021rkz,Bautista:2021wfy,Bernuzzi:2008rq} to describe dS Love numbers.\\ 
    
    Following this, we focus on the microscopic details of the compact object \cite{Kol:2011vg, Creci:2021rkz,Hui:2020xxx} to compute the macroscopic scattering coefficients of the worldline EFT. At the microscopic level, we will solve the scalar field equations perturbatively on the SdS background with the ingoing boundary condition at the black hole horizon.\\

    Since the dS worldline EFT was defined on the conformally flat coordinates of the Poincaré patch, we will use the flat slicing coordinates \cite{1981PTPh,Shiromizu:2001bg, Cosmologicalflat} of the SdS black hole to match the microscopic picture with the macroscopic worldline EFT \cite{Kol:2011vg,Creci:2021rkz}. However, since the location of the horizon is more apparent in the static chart of the SdS black hole, the scalar field equation will be solved in the static chart. Following this, we will perform a coordinate transformation from the static coordinates to the flat slicing coordinates within the static patch of the SdS black hole, allowing for matching with the dS worldline EFT.\\ 
    
    Computing the scattering coefficients from the microscopic picture requires a near zone - far zone matching calculation of the kind discussed in \cite{Creci:2021rkz,Cardoso:2019nis,Cardoso:2004nk,Castro:2010fd}. In this work, we will perform the calculation with two different notions of the near zone. The first is an extension of the conventional notion used for Schwarzchild black holes \cite{Creci:2021rkz,Cardoso:2019nis,Cardoso:2004nk}, with the consideration of an additional length scale related to the cosmological constant. The second is a notion of proximity to the horizon developed through a perturbative expansion in $(R-R_\mathrm{h})/R_\mathrm{h}$, $R_\mathrm{h}$ being the black hole horizon; this notion of proximity can best be described as the near horizon region \cite{Castro:2010fd}. In this work, we have used both these notions of the near zone to illustrate the intricacies associated with the appropriate notion of a near zone in Love number computations.\\
    
    We will start with a brief review of Love numbers in \ref{Love numbers}. Following this, we have the two main parts of this work: \ref{deEFT} and \ref{Love numbers of Schwarzschild de Sitter}. In \ref{deEFT}, we will extend the notion of the worldline EFT developed for asymptotically flat compact objects to the asymptotically non-flat case. Here, we shall illustrate how the worldline EFT can be used to define Love numbers for asymptotically non-flat spacetimes. Following this, we will specialise to asymptotically de Sitter compact objects. We will work in the conformally flat coordinates on the Poincaré patch of the dS spacetime to simplify the calculations. Finally, we obtain an expression for the scalar Love numbers as measured by a faraway comoving observer on a dS background. Here, we shall make some key observations regarding the response function for the compact object being time-dependent due to the observer time not being Killing.\\
    
    In \ref{Love numbers of Schwarzschild de Sitter}, we use the worldline EFT developed in \ref{deEFT} and the associated notion of Love numbers to obtain an explicit expression for the SdS black hole Love numbers up to $\mathcal{O}(\Lambda M^2)$. This shall proceed through solving the scalar field equation on a SdS background perturbatively in $\Lambda M^2$ with the ingoing boundary condition at the horizon. Here, we will perform a near zone - far zone matching calculation to extract the scattering coefficients for a SdS black hole. Following this, we have a summary and discussion of the results in \ref{Summary and Discussion} and the conclusions in \ref{Conclusion}.\\ 
    
    \textit{Notations and Conventions:} Throughout the paper, we have used the mostly positive signature convention. The Greek indexing runs over both spatial and temporal directions. The Roman indexing is restricted to spatial directions. $L =(i_1 , i_2\cdots , i_\ell )$ is a multi-index and each $i$ runs over the spatial directions $\{1,2,3\}$. $A_L$ represents a spatial symmetric traceless tensor with the spatial multi-index $L$. For example the unit vector $n^{L=2}=n^{ij}=n^in^j-\frac{1}{3}\delta^{ij}$. We have also set the fundamental constants $G$ and $c$ to unity.\\

    \section{Tidal Love numbers}
    \label{Love numbers}
    
    Let us consider a mass distribution $\rho(\Vec{x})$ in Newtonian gravity. An external massive body interacts gravitationally with $\rho(\Vec{x})$ such that it induces a change $\delta\rho(\Vec{x})$ in mass distribution resulting in an additional multipole moment $Q_L$. Then the gravitational potential sufficiently away from the centre of mass of the mass distribution $\rho(\Vec{x})$ will be given by \cite{Poisson(Book):2014}
    \begin{equation}
    \begin{aligned}
                U_{tot}(r)=&~U_\rho(r)-\sum_{\ell=2}^\infty\frac{(\ell-2)!}{\ell!}{n^L\mathcal{E}_L~r^\ell}\\     &\hspace{2.5cm}+\sum_{\ell=2}^\infty\frac{(2\ell-1)!!}{\ell!}\frac{n^LQ_L}{r^{\ell+1}}.
    \end{aligned}
    \end{equation}
    Here we have $\mathcal{E}_L=-\frac{1}{(\ell-2)!}\partial_L U_{\mathrm{ext}}(t)\big|_{\Vec{x}=0}$, the tidal field exerted by the external body. $U_{\mathrm{ext}}$ is the gravitational potential exerted by the external body and $U_\rho$ is the potential sourced by the unperturbed mass distribution $\rho(\Vec{x})$. \\
    
    We can further note that the change in the mass distribution $\delta\rho(\Vec{x})$ in response to an external field depends on the properties of the matter and can be quantified through the associated change in the multipole moment in response to the external tidal field. In fact, for spherically symmetric systems, we may write
    \begin{equation}
        Q_L(t)=k_\ell \mathcal{E}_L(t)-\tau_0 \nu_\ell\dot{\mathcal{E}}_L(t)+\cdots~.
    \end{equation}
    In the above equation, we refer to $k_\ell$ as the Love number of the mass distribution and $\tau_0$ is a time scale characterising the change in the mass distribution in response to the time variation of the tidal field. $\nu_\ell$ represents the loss of energy due to tidal heating, which is called the tidal dissipation numbers. The ellipses represent possible higher-order dependence on the time variation of $\mathcal{E}_L$. Performing a Fourier transformation on the above equation, we can see that 
    \begin{equation}
       \begin{gathered}
            Q_L(\omega)=-F_\ell(\omega)\mathcal{E}_L(\omega).
       \end{gathered}
    \end{equation}
    Where $F_\ell(\omega)$ is called the tidal response coefficient, its real part contains information regarding the tidal response, and the imaginary part quantifies the tidal dissipation  \cite{Poisson(Book):2014}.\\

    Earlier works have extended the notion of tidal Love numbers of compact objects to relativistic systems. The definition relied on the identification of the coefficients characterising the growing and decaying part of the perturbing field on the background metric at asymptotic infinity for asymptotically flat spacetimes \cite{Zhang:1986cpa,Hinderer:2007mb,Binnington:2009bb}. It has also been shown that one can define Love numbers, as observed by a distant free-falling observer in terms of the scattering coefficients of the perturbation, within the framework of a world line EFT for asymptotically flat spacetimes \cite{Creci:2021rkz,Bautista:2021wfy,Bernuzzi:2008rq}.\\

    Despite the significant progress made in the field of Love numbers for asymptotically flat compact objects within general relativity \cite{Damour:2009vw,Kol:2011vg,Chia:2021,Ivanov:2022qqt,Pani:2015hfa,Landry:2015zfa,Hui:2020xxx,Creci:2021rkz,Addressingissues:2023,Ivanov:2022hlo} and non-vacuum GR theories of gravity \cite{Cardoso:2017cfl,Nair:2022xfm,Katagiri:2023yzm,Katagiri:2023umb,DeLuca:2022tkm,Chakraborty:2023zed}, there is very little literature on the Love numbers of asymptotically non-flat compact objects. We note that a major challenge regarding this concerns the notion of Love numbers by identifying the response and source terms at asymptotic infinity as the decaying and growing parts of the perturbed field, respectively, cannot be extended straightforwardly to non-flat spacetimes. However, the notion of Love numbers within the framework of a worldline EFT can be extended to non-flat spacetimes, where the Love numbers can be defined in terms of the scattering coefficients of the perturbing field as measured by a distant observer. \\ 

    In the following section, we shall develop a notion of Love numbers for comoving compact objects on a de Sitter background as measured by a comoving observer within the framework of a worldline EFT on a de Sitter background. We shall restrict ourselves to the Love number for an arbitrary scalar field to simplify the computation. Along the way, we will also list sufficient conditions for non-flat spacetimes for which a definition of Love numbers similar to ours is possible. In \ref{Love numbers of Schwarzschild de Sitter}, we shall use the notion of Love numbers developed in \ref{deEFT} to compute the SdS scalar Love numbers observed by a distant comoving observer.
    
    \section{Worldline Effective Field Theory for de Sitter}
    \label{deEFT}
    
    This section will construct a worldline EFT for compact objects on a non-flat background. We will note down certain features of the background spacetime, which allows for a definition of scalar Love numbers using worldline EFT. Then we specialize to a comoving compact object on a de Sitter (dS) spacetime. We will use the worldline EFT to define the scalar Love numbers for the compact object through the scattering coefficients of the scalar field as measured by a comoving observer far away from the compact object.
    
\subsection{The Setup}
\label{The Setup}
We will consider a compact object sufficiently far away from the observer on a dS background such that the characteristic length associated with the compact object $r_0$ is much smaller than the coordinate separation between the compact object and the observer. Such a compact object may be modelled as a point particle moving along its worldline after integrating out its characteristic length scale $r_0$. The finite size effects of the compact object will be accounted for through the presence of extra field couplings on the point particle worldline, within the framework of a worldline EFT \cite{Goldberger:2004jt, Kol:2007rx, Kol:2011vg, Chakrabarti:2013xza, Hui:2020xxx, Creci:2021rkz}.
    
The compact object can interact with the scalar field on the dS background through finite size interactions. In particular, the tidal field $\mathcal{E}_L=\nabla_L\phi$, generated by the scalar field, can deform the compact object and give rise to the multipole moments $Q^L$. For such a system, the effective action will be of the form \cite{Kol:2007rx,Kol:2011vg,Creci:2021rkz,Goldberger:2004jt};    
\begin{equation}    
\label{totalact}
S_{\mathrm{total}}=S_{\mathrm{pp}}+S_{\phi}+S_{\mathrm{int}}+S_{\mathrm{G}}+S_{\mathrm{tidal}}~.
\end{equation}
Here, $S_{\rm pp}$ is the point particle action, $S_{\phi}$ is the action for the scalar field and $S_{\rm G}$ is the gravitational action, each of which can be given by, 
\begin{align}        
S_{\mathrm{pp}}&=-M \int d \tau \sqrt{-u_\mu u^\mu}~,
\\
S_{\phi}&=-\frac{K_\phi}{2} \int d^4 x \sqrt{-g}~\nabla^\mu \phi \nabla_\mu \phi~,
\\
S_{\mathrm{G}}&=\frac{1}{16 \pi} \int d^4 x \sqrt{-g} (R-2\Lambda)~.       
\end{align}
We have assumed the mass of the point particle to be $M$, which is moving along its worldline with affine parameter $\tau$; $K_{\phi}$ scales the overall stregth of the scalar field, and we have written the gravitational action with a positive cosmological constant, as our interest lies in the asymptotically de Sitter spacetimes. Among other terms in \ref{totalact}, $S_{\mathrm{int}}$ is the action describing the internal dynamics of the finite size effects of the compact object \cite{Chakrabarti:2013xza} and $S_{\mathrm{tidal}}$ is the part of the action describing the interaction between the tidal effect of the scalar field and the multipole moments of the compact object. In the analysis that follows, we will not bother about the complicated internal dynamics of the compact object, which are contained in $S_{\mathrm{int}}$ \cite{Chakrabarti:2013xza}; instead, we aim to infer $Q^L$ from the scattering coefficients of the scalar field as observed by a distant observer. Thus for our purpose providing an expression for the action $S_\mathrm{tidal}$ sufficies, which reads,
\begin{align}
\label{tidact}
S_{\mathrm{tidal}}&=\int \sqrt{-g} d^4 x\Bigg[-K_{\mathrm{T}}\int d \tau \sqrt{-u_\mu u^\mu}\frac{1}{\sqrt{-g}}
\nonumber
\\
&\times\delta^{(4)}(x^\mu-z^\mu(\tau))\sum_{\ell=0}^{\infty} \frac{1}{\ell !} Q^L(z^\mu(\tau)) \nabla_L\phi\Bigg]~.
\end{align}
Here, $Q^L$ are the multipole moments of the compact object; $\nabla_L \phi$ are the scalar tidal fields ($\mathcal{E}_L$); $u^\mu$ is the 4-velocity of the object; $z^\mu(\tau)$ is the worldline of the compact object and $K_{\mathrm{T}}$ is the coupling constant characterising the interaction between scalar field and the compact object. The delta function $\delta^{(4)}(x^\mu-z^\mu(\tau))$ ensures that the scalar field only interacts with the compact object, whose location in the spacetime is given by $z^\mu(\tau)$, within the framework of the worldline EFT. In what follows, we will use the worldline EFT approach to define the Love numbers for compact objects which are not asymptotically flat. 
    
\subsection{Love numbers from worldline EFT}
\label{Computing the Love number for asymptotically de Sitter compact objects.}
Our goal in this section is to solve the scalar field equation on the dS background and identify the two linearly independent parts of the solution, along with their constant coefficients. Following this, we will determine the multipole moment $Q^L$, in terms of these constant coefficients, which can be associated with the scattering coefficients observed by a distant observer \cite{Creci:2021rkz}. Similarly, the tidal field $\mathcal{E}_L$ can also be determined in terms of these constant coefficients, and hence the response function $F_\ell(t)$ can be determined, whose real part gives the Love numbers, $k_\ell(t)$ \cite{Poisson(Book):2014, Kol:2011vg, Binnington:2009bb, Creci:2021rkz, Addressingissues:2023}. Note that the response function can explicitly be a function of time if the spacetime is not static. Thus we obtain,
\begin{align}
\label{reponsf}
Q_L(t)=-F_\ell(t)\mathcal{E}_L(t)~;
\qquad
k_\ell(t)\equiv \frac{1}{2}\mathrm{Re}\left[F_\ell(t)\right]~.
\end{align}
Note that the tidal field $\mathcal{E}_L$ should be understood as the finite part of $\nabla_L\phi$, while the $Q_{L}$ arises from the divergent part of $\nabla_L\phi$, at the origin, in some appropriate radial coordinate, evaluated on the world line of the body \cite{Creci:2021rkz}.
    
As we are interested in extracting the scalar Love numbers from the scattering of the scalar field from the compact object, we will consider the Euler-Lagrange equation for the scalar field, which gives
\begin{align}
\label{phil}
\square\phi&=\sum_{\ell=0}^{\infty}T_{\phi}^{\ell}~, 
\\
T^{\ell}_{\phi}&=\frac{K_\mathrm{T}}{K_\phi} \frac{(-1)^{\ell}}{\ell !}  \int d \tau \nabla_L\Bigg[\frac{\sqrt{-u_\mu u^\mu}}{\sqrt{-g}}
\nonumber
\\ 
&\qquad\times Q^L(z^\mu(\tau)) \delta^{(4)}\left(x^\mu-z^\mu(\tau)\right)\Bigg]~, 
\label{phil2}
\end{align}
where, $\square=\nabla_\mu\nabla^\mu$ and $K_\phi$ is the coupling constant characterising the scale field action.

In order to find solutions to the above differential equation, we express the scalar field $\phi$ as a sum of various angular modes $\phi_\ell$, such that,
\begin{equation}
\phi=\sum_{\ell=0}^\infty\phi_\ell~.
\end{equation}
Plugging the above decomposition in \ref{phil} will imply that each $\phi_\ell$ will satisfy the equation $\square \phi_{\ell}=T^{\ell}_{\phi}$, where $T^{\ell}_{\phi}$ has already been defined in \ref{phil2}.

\subsubsection{Defining Love numbers using worldline EFT}
\label{Defining Love numbers using worldline EFT.}
    
Upon close examination of \ref{phil}, it is apparent that if $\phi_0$ is a solution to the $\ell=0$ differential equation, then the solution to the $\ell$th mode may be given by $\nabla_L \phi_0$ if $[\square,\nabla_L]=0$. This means that if we can write the background spacetime in a chart, where the above commutation holds true, we may find $\phi_\ell$ by simply solving for $\phi_0$. Further, if the chart has the properties $\nabla_L \phi=\partial_L \phi$ and $\nabla_L\sqrt{-g}=0$, we can make use of the results in \cite{destrider,eqbook,Blanchet:1985sp} and perform an analysis similar to the flat space worldline EFT \cite{Creci:2021rkz} to get an explicit expression for the Love number in terms of the scattering coefficients of the scalar field as observed by a far away observer. The above observations imply that for any spacetime that meets the above-specified conditions within some chart, we can provide a definition of Love number for compact objects within the framework of a worldline EFT, which we are going to illustrate for the specific case of compact objects in asymptotically de Sitter spacetimes.
    
\subsubsection{de Sitter universe in the Poincar\'{e} patch}
\label{de Sitter in the Poincaré patch}
    
Motivated by the above discussion, we will choose to work in the Poincaré patch of the de Sitter spacetime employing the conformal coordinates, where the metric can be expressed as;    
\begin{equation}
\label{de Sitterconf}
dS^2=c(\eta)^2\left[-d \eta^2+d \vec{x}^2\right]~;
\qquad 
c(\eta)=-\frac{1}{H \eta}~.
\end{equation}
Here $H\equiv\sqrt{\Lambda/3}$, where $\Lambda$ is the positive cosmological constant associated with the de Sitter universe \cite{Spradlin:2001pw, Akhmedov:2013vka, Ashtekar:2015lxa, Rajeev:2019okd, Chakraborty:2021ezq, Gaur:2023hmk}. Working in the Poincar\'{e} patch in the conformal coordinates, we can observe that for any scalar $S(\eta,x^i)$ the following identies hold:
\begin{enumerate}
\label{properties}
\item $\left[\nabla_L,\square\right]S(\eta,x^i)=0$~,
\item $\nabla_L S(\eta,x^i)=\partial_L S(\eta,x^i)$~,
\item  $\nabla_L\sqrt{-g}=0$~.
\end{enumerate}
As a consequence we can construct the $\phi_\ell$ from $\phi_0$ as: $\phi_\ell=\nabla_L\phi_0$. Further the above identities also implies that $\nabla_L\phi_0=\partial_L\phi_0$; and the source term, $T_\phi^\ell$ of \ref{phil} can be shown to be proportional to the derivatives of the 3-dimensional delta function.

By considering a family of point particles moving with the cosmic flow such that their coordinates can be found as $z^\mu(\tau)=\left(z^0(\tau),0,0,0\right)$, while satisfying the normalization condition $u^\mu u_\mu=-1$, the source term $T^\ell_\phi$ can be further simplified to
\begin{align}
\label{sorceconf}
T^\ell_\phi&=\frac{K_\mathrm{T}}{K_\phi} \int d \tau \frac{(-1)^{\ell}}{\ell!} \nabla_L \Bigg[\frac{\delta^{3}\left(x^i\right)}{\sqrt{-g}} Q^L(z^\mu(\tau))
\nonumber
\\
&\qquad\qquad \times\delta\left(\eta-z^0(\tau)\right)\Bigg]
\nonumber
\\             
&=\frac{(-1)^{\ell}K_\mathrm{T}}{\ell!\sqrt{-g}K_\phi} Q^L(\eta)\,\partial_L\left[\delta^{3}\left(x^i\right)\right]\times\frac{d\tau}{d\eta}~.
\end{align}
Subsequent computation of the $\ell=0$ mode and, later, the determination of the higher $\ell$ modes, which are performed by the action of $\nabla_L$ on $\phi_{0}$, will require use of the properties (1) --- (3), listed above. 

\subsubsection{Obtaining the zero mode solution}
\label{Solving for the zero mode solution}

We can find the zero mode $\phi_0$ by solving \ref{phil} with $\ell=0$, which reduces the source term to zero. Following this, and the symmetries of the de Sitter universe in the Poincar\'{e} patch, we consider the ansatz: $\phi_0=w(\eta,x^i)/c(\eta)$; with $c(\eta)$ being the scale factor of de Sitter universe, as defined in \ref{de Sitterconf}. Substituting the above ansatz in \ref{phil} with $\ell=0$, we obtain the following differential equation for $w(\eta,x^i)$
\begin{equation}
\begin{gathered}
\frac{\partial^2 w}{\partial \eta^2}-\nabla^2 w-\frac{2w}{\eta^2}=0~,
\end{gathered}
\end{equation}
As we are solving for the $\ell=0$ mode, there is no angular dependence in $w$, and hence we may express $w(\eta,\textbf{x})$ as, $w(\eta,\textbf{x})=v(r)u(\eta)$. Since space and time sectors do not talk to each other, it follows that $v(r)$ must satisfy the equation $\nabla^2 v(r)=-\Omega^2 v(r)$, where $\Omega$ is a constant and $u(\eta)$ satisfies the following differential equation
\begin{equation}
\frac{\partial^2 u}{\partial \eta^2}+\left(\Omega^2-\frac{2}{\eta^2}\right)u=0~.
\end{equation}
The above differential equation can be solved by using a linear combinations of Hankel functions \cite{Akhmedov:2013vka}, and $u(\eta)$ takes the following form,
\begin{equation}
u(\eta)=\sqrt{\eta}\left(A H^{(1)}_{\frac{3}{2}}(\Omega\eta)+B H^{(2)}_{\frac{3}{2}}(\Omega\eta)\right)~.
\end{equation}
From the properties of the Hankel function, it follows that $H^{(1)}_{\alpha}(z)\sim (1/\sqrt{z})e^{iz}$ and $H^{(2)}_{\alpha}(z)\sim (1/\sqrt{z})e^{-iz}$, for $|z|\to \infty$. Furthermore, as in the flat spacetime, here also we impose the condition that the zero mode should behave as $e^{i\omega \eta}$ for $\eta\to -\infty$, which when coupled with the above properties of the Hankel function, demands $A=1$, and $B=0$.

The spatial sector, on the other hand, satisfies the equation $\nabla^2 v(r)=-\Omega^2 v(r)$, which can also be solved by Hankel functions, if we expand the Laplacian in the spherical polar coordinates. Therefore, the zero mode $\phi_0$ on the dS background takes the form
\begin{align}
\label{phi0}
\phi_0(r,\eta)&=\frac{\eta^{3/2}}{\sqrt{r}}\sqrt{\frac{\pi\Omega}{2}}H^{(1)}_{\frac{3}{2}}(\Omega\eta)
\nonumber
\\
&\times \Bigg\{C_{\mathrm{in}}e^{i\frac{\pi}{2}}H^{(1)}_{\frac{1}{2}}(\Omega r)
+C_{\mathrm{out}}e^{-i\frac{\pi}{2}}H^{(2)}_{\frac{1}{2}}(\Omega r)\Bigg\}~,
\end{align}
where $C_{\mathrm{in/out}}$ are the ingoing and outgoing scattering coefficients for the scalar field as observed by a distant comoving observer. The extra factors involving $(\sqrt{\pi\Omega/2})e^{\pm i\pi/2}$ have been introduced to ensure the appropriate ingoing and outgoing behaviours of the Hankel function at large $r$ \cite{Creci:2021rkz}. However, both the Hankel functions are ill-behaved near the origin $r=0$, where the compact object is placed, and hence we would like to modify the Hankel functions to the Bessel functions, such that at least one of the solution is finite at the location of the compact object.

\subsubsection{Change of Basis}
\label{Basis Change}
We will next perform a basis change from the Hankel functions to the Bessel functions for the spatial part of \ref{phi0}, to ensure regularity for at least one of the solutions at the location of the compact object. This is achieved through the following equations \cite{eqbook},
\begin{align}
J_\mathrm{p}(\Omega r)&=\frac{1}{2}\left(H_\mathrm{p}^{(1)}(\Omega r)+H_\mathrm{p}^{(2)}(\Omega r)\right)~,
\\
Y_{p}(\Omega r)&=\frac{1}{2 i}\left(H_\mathrm{p}^{(1)}(\Omega r)-H_\mathrm{p}^{(2)}(\Omega r)\right)~.
\end{align}
such that the zero mode solution, as in \ref{phi0}, can be expressed as,  
\begin{align}
\label{phi0bassel}
\phi_0(r,\eta)&=\frac{\eta^{3/2}}{\sqrt{r}}\sqrt{2\pi \Omega}\,H^{(1)}_{\frac{3}{2}}(\Omega\eta)
\nonumber
\\ 
&\times \Bigg\{C_{\mathrm{irr}}Y_{\frac{1}{2}}(\Omega r)+C_{\mathrm{reg}}J_{\frac{1}{2}}(\Omega r)\Bigg\}~,
\end{align}
where, the arbitrary constants $C_{\rm reg}$ and $C_{\rm irr}$ can be expressed in terms of $C_{1}$ and $C_{2}$ as,
\begin{equation}
\begin{gathered}
C_{\mathrm{reg}}=i~\frac{(C_{\mathrm{in}}-C_{\mathrm{out}}) }{2}~;
\quad  
C_{\mathrm{irr}}=-~\frac{(C_{\mathrm{in}}+C_{\mathrm{out}})}{2}~. 
\end{gathered}
\end{equation}
Note that in \ref{phi0bassel}, the term $J_{\frac{1}{2}}(\Omega r)$ is regular at $r=0$, while the other term, namely $Y_{\frac{1}{2}}(\Omega r)$ is irregular at $r=0$. With this change of basis, we now wish to compute the multipole moments of the compact object and relate it to the tidal field to determine the tidal Love number in terms of the coefficients $C_{\rm irr}$ and $C_{\rm reg}$.  

\subsubsection{Relating the multipole moments and tidal fields}
\label{Finding the multipole moments.}
In this section we will first compute the $\ell$th order moment of the scalar field  $\phi_\ell$ using the simple relation: $\phi_\ell=\partial_L\phi_0$, for each independent solutions of $\phi_{0}$. Following this, we will evaluate $\square\phi_\ell$ where the spatial part of the $\square$ operator is interpreted as a distributional derivative \cite{Creci:2021rkz, destrider}. Note that as we are working in the Poincaré patch of the de Sitter spacetime, which is conformally flat, all the above computations may proceed in a manner similar to that of the flat space \cite{Creci:2021rkz}. This results in the following expression for $\phi_\ell$,    
\begin{align}
\label{phil1}
\phi_{\ell}&=\eta^{3/2}\sqrt{2 \pi \Omega}~H^{(1)}_{\frac{3}{2}}(\Omega \eta) \Bigg\{C^L_{\mathrm{reg}} ~\partial_L\left( r^{-1/2} J_{\frac{1}{2}}(\Omega r)\right)
\nonumber
\\            
&\qquad +C^L_{\mathrm{irr}}\sqrt{2 \pi \Omega}~\partial_L\left( r^{-1/2} Y_{\frac{1}{2}}(\Omega r)\right)\Bigg\}~,
\end{align}
where $C^L_{\mathrm{reg/irr}}$ are arbitrary constants characterising the regular and irregular part of the solution to the second order differential equation in \ref{phil}. We can further simplify \ref{phil1} using the following identities \cite{Blanchet:1985sp},
\begin{align}
\partial_L g(r)&=n_L r^{\ell}\left(\frac{1}{r} \frac{\partial}{\partial r}\right)^{\ell} g(r)~,
\nonumber
\\
\left(\frac{1}{z} \frac{d}{d z}\right)^k\left(z^\nu \mathcal{B}_\nu(z)\right)&=z^{\nu-k} \mathcal{B}_{\nu-k}(z)~, 
\nonumber
\\ 
\left(\frac{1}{z} \frac{d}{d z}\right)^k\left(z^{-\nu} \mathcal{B}_\nu(z)\right)&=(-1)^k z^{-\nu-k} \mathcal{B}_{\nu+k}(z)~,
\end{align}
where we have $g(r)$ to be some arbitrary function of $r$, and $\mathcal{B}_\nu(z)$ representing the Bessel functions, either $J_\nu(z)$ or $Y_\nu(z)$. This allows to simplify \ref{phil1} to read \cite{Creci:2021rkz},
\begin{align}
\label{philirrreg}
\phi_\ell=&\frac{\eta^{3/2}}{\sqrt{r}}\sqrt{2 \pi \Omega}~\Omega^{\ell} (-1)^{\ell}~H^{(1)}_{\frac{3}{2}}(\Omega \eta)
\nonumber
\\
&\times n_L\left[C_{\mathrm{reg}}^L J_{1/ 2+\ell}(\Omega r)+C_{\mathrm{irr}}^L Y_{1 / 2+\ell}(\Omega r)\right]~.
\end{align}
Having determined $\phi_\ell$, our next goal is to obtain an expression for the quadrupole moment $Q^L(\eta)$ in terms of the arbitrary constants $C_{\mathrm{reg}}^L$ \& $C_{\mathrm{irr}}^L$. For this purpose, we will substitute \ref{philirrreg} in \ref{phil}, and subsequent comparison with \ref{sorceconf} yields,
\begin{align}
\label{multiopoll}
Q^L(\eta)&=\frac{(-1)^{\ell} \ell !}{H^2\sqrt{\eta}}H^{(1)}_{\frac{3}{2}}(\Omega \eta)~\frac{8 \pi K_\phi}{K_\mathrm{T}}C_{\mathrm{irr}}^L\times\frac{d\eta}{d\tau}~.
\end{align}
The tidal part can be determined using the finite part of \ref{reponsf}, i.e., by evaluating $\phi_{\ell}$ on the worldline of the compact object, obtained by taking the $r\rightarrow0$ limit of $\nabla_L\phi$  \cite{Creci:2021rkz,destrider}, resulting in,
\begin{equation}
\label{multipoleandtidal}
\mathcal{E}_L(\eta)=\sqrt{\pi}\,\eta^{3/2}\,H^{(1)}_{\frac{3}{2}}(\Omega \eta) \frac{(-1)^{\ell} 2^{\ell+1}\ell!}{\Gamma\left(\frac{1}{2}+\ell+1\right)} C_{\mathrm{reg}}^L\left(\frac{\Omega}{2}\right)^{1+2 \ell}~.
\end{equation}
Using \ref{de Sitterconf} to lower the spatial indices of the multipole moment tensor $Q^{L}$ in \ref{multiopoll}, we obtain
\begin{equation}
Q_L(\eta)=\frac{(-1)^{\ell}\ell!}{(H\eta)^{2(\ell+1)}}\eta^{3/2}H^{(1)}_{\frac{3}{2}}(\Omega \eta)\frac{8 \pi K_\phi}{K_\mathrm{T}}C_{\mathrm{irr}}^L\times\frac{d\eta}{d\tau}~. 
\end{equation}
So far, we have kept the proper time of the particle unspecified, however our computations are done for a point particle comoving with the Hubble flow, which means that its proper time will be the cosmological time $t$, related to the conformal time through the relation: $e^{Ht}=-(1/H\eta)$. Therefore, in terms of the cosmological time of the comoving observer, the multipole moments can be expressed as\footnote{ Looking at the multipole and the tidal field described in \ref{multiopoll}, \ref{multipoleandtidal} and \ref{mutipolcosmo}, we may note that it lacks a factor $\sqrt{2\pi}$, in relation to the flat space expressions, this is a result of the convention of the basis expansion used.},
\begin{align}
\label{mutipolcosmo}
Q_{L}(t)&=(-1)^{\ell}\ell!~ \frac{8 \pi K_\phi}{K_\mathrm{T}}C_{\mathrm{irr}}^L\times e^{H t(2\ell+1)}
\nonumber
\\
&\qquad \times\Big[\eta(t)^{3/2}H^{(1)}_{\frac{3}{2}}(\Omega \eta(t))\Big]~.
\end{align}
Thus the computation of the response function in terms of cosmological time follows by taking the ratio of the multipole moment in \ref{mutipolcosmo} with the tidal field in \ref{multipoleandtidal}, resulting in
\begin{align}
\label{response}
F_\ell(t)&=-\frac{K_\phi}{K_\mathrm{T}}\frac{4 \pi^{1 / 2}}{2^{\ell}}\left(\frac{2}{\Omega}\right)^{1+2 \ell} \Gamma\left(\frac{2\ell+3}{2}\right)\frac{C^L_{\mathrm{irr}}}{C^L_{\mathrm{reg}}}~e^{H t(2\ell+1)}
\nonumber
\\
&\equiv \frac{K_\phi}{K_\mathrm{T}}\Tilde{F}_\ell(t)~.
\end{align}
Here we have identified the normalised response function of the object to be $\Tilde{F}_\ell(t)$, which reads,
\begin{equation}
\label{responsereg}
\Tilde{F}_\ell(t)=-\frac{4 \pi^{1 / 2}}{2^{\ell}}\left(\frac{2}{\Omega}\right)^{1+2 \ell} \Gamma\left(\frac{2\ell+3}{2}\right)\frac{C^L_{\mathrm{irr}}}{C^L_{\mathrm{reg}}}\times e^{H t(2\ell+1)}~.
\end{equation}
From the microscopic perspective, the ratio $C^L_{\mathrm{irr}}/C^L_{\mathrm{reg}}$ depends on the nature of the compact object. From the macroscopic worldline EFT perspective, this ratio can be associated with the ratio of the ingoing and outgoing scattering coefficients.

Some comments are in order regarding the multipole moment and the Love numbers being a function of time. This is a consequence of the background spacetime not being stationary. As a result, the temporal part of the field cannot be expressed in the Fourier basis; instead, we have to use the Hankel basis. A similar time dependence on the gravitational multipole moments on a de Sitter background was reported earlier in \cite{Chakraborty:2021ezq}.
    
To explicitly see the relation of  $C^L_{\mathrm{irr}}/C^L_{\mathrm{reg}}$ with the scattering coefficients we may go to the Hankel basis on the spatial sector of \ref{philirrreg} and using the asymptotic behaviour of the Hankel function \cite{eqbook}  to identify the ingoing and outgoing coefficients \cite{Creci:2021rkz}. This will allow us to express $C^L_\mathrm{reg/irr}$ as follows;
\begin{equation}
\label{hanktobas}
\begin{aligned}
& C_{\mathrm{reg}}^L=\frac{(-1)^{\ell}}{2} i^{\ell}\Bigg[C_{\mathrm{in}}^L e^{i \frac{\pi}{2}(\ell+1)}+(-1)^{\ell} C_{\mathrm{out}}^L e^{-i \frac{\pi}{2}( \ell+1)}\Bigg], \\
& C_{\mathrm{irr}}^L=\frac{(-1)^{\ell}}{2} i^{\ell+1}\Bigg[C_{\mathrm{in}}^L e^{i \frac{\pi}{4}( \ell+1)}\\
&\text{\hspace{3.25cm}}+(-1)^{\ell+1} C_{\mathrm{out}}^L e^{-i \frac{\pi}{4}(\ell+1)}\Bigg], \\
\end{aligned}
\end{equation}
where $C^L_\mathrm{in/out}$ are the scattering coefficients of the scalar field as observed by a distant comoving observer. Using the above relations and \ref{response}, we can define the response function of a comoving compact object on a de Sitter background on the Poincaré patch in terms of the ingoing and outgoing scattering coefficients observed by a faraway comoving observer within the framework of de Sitter worldline EFT.
    
Despite this section mostly focusing on using worldline EFT for asymptotically de Sitter compact objects, a similar analysis is possible for spacetimes having features discussed in \ref{Defining Love numbers using worldline EFT.}. Our next goal would be to perform a microscopic calculation to compute $C^L_{\mathrm{irr}}/C^L_{\mathrm{reg}}$ for small SdS black holes in order to get an expression for SdS black hole scalar Love numbers.

    \section{Love numbers of Schwarzschild de Sitter}
    \label{Love numbers of Schwarzschild de Sitter}
    
    In this section, we will compute the scalar Love numbers for a Schwarzschild de Sitter (SdS) black hole. The calculations presented here will be at the microscopic level, where we will compute $C^L_{\mathrm{irr}}/C^L_{\mathrm{reg}}$ for a small SdS black hole. This ratio can further be associated with the scattering coefficients observed by a distant comoving observer, through \ref{hanktobas}.\\
    
    The computation involves us exploring the consequences of the near horizon physics on the asymptotic behaviour of the scalar fields in the static chart of the SdS black hole, then performing a coordinate transformation on the static patch from the static coordinate to the flat slicing coordinates, and finally matching the asymptotic behaviour with the macroscopic background dS worldline EFT. This matching will allow us to express the Love numbers for a SdS black hole as observed by a distant comoving observer. We will also give a functional definition for the tidal response coefficient for asymptotically de Sitter compact objects based on computation in the static coordinates.
    
    \subsection{Flat slicing coordinates for SdS and the worldline EFT on the Poincaré patch of dS.}
    \label{Flat slicing coordinates for SdS and the EFT on the Poincaré patch of dS}
    
    If we consider spherically symmetric compact objects, whose exterior can be written in the form \cite{1981PTPh,Shiromizu:2001bg,Cosmologicalflat} (known as the flat slicing for SdS black holes);
    \begin{equation}
    \label{confbh}
    \begin{aligned}
        & d s^2=-g(r,\eta)d \eta^2+h(r,\eta)d \vec{x}^2~, \\
        & g(r, \eta)=a^2(\eta)\left[1-\frac{M}{2 a(\eta) r}\right]^2\left[1+\frac{M}{2 a(\eta) r}\right]^{-2},\\
        & h(r,\eta)=a^2(\eta)\left[1+\frac{M}{2 a(\eta) r}\right]^4,\\
        &a(\eta)=-1/H\eta~.
    \end{aligned}
    \end{equation}
    In the large $r$ limit, the above metric becomes the dS spacetime in the Poincaré patch. So in the same spirit as the microscopic calculations in the flat spacetime \cite{Kol:2011vg, Creci:2021rkz,Hui:2020xxx}, we should be matching the scalar field in the large $r$ limit of the flat slicing coordinates of the SdS black hole with the scalar field on the dS worldline EFT.\\
    
    The angular part of the scalar field equation in the large $r$ limit of a spacetime given by the metric in \ref{confbh} becomes separable in the spherical harmonic basis, $\mathrm{Y}_{\ell m}$ and  the solution can be seen to be of the form
    \begin{equation}
    \label{philflatfar}
        \begin{gathered}
        \begin{aligned}
            \phi_\ell=&\eta\sqrt{\eta} H^{(1)}_{\frac{3}{2}}(\Omega \eta) \sqrt{2 \pi \Omega}~ \Omega^{\ell} (-1)^{\ell}~r^{-1 / 2} \\
            & \times \Bigg\{ {}^{\mathrm{flat}}A_{\mathrm{reg}}^{\ell } J_{1/ 2+\ell}(\Omega r)+{}^{\mathrm{flat}}A_{\mathrm{irr}}^{\ell } Y_{1 / 2+\ell}(\Omega r)\Bigg\}\\
        \end{aligned}
        \end{gathered}
    \end{equation}
    where we have
    \begin{equation}
    \label{const1}
        {}^{\mathrm{flat}}A_{\mathrm{reg/irr}}^{\ell }=\sum\limits_{m=-\ell}^\ell{}^{\mathrm{flat}}A_{\mathrm{reg/irr}}^{\ell m}\mathrm{Y}_{\ell m}(\theta,\phi) 
    \end{equation}
    with ${}^{\mathrm{flat}}A_{\mathrm{reg/irr}}^{\ell m}$ being arbitrary coefficients associated with the basis expansion for each $m$. Further, we may choose the parameter $\Omega$ to be the same as in \ref{philirrreg}.\\
    
    We can re-express the above equation in terms of the unit vectors ($n_L$) using
   \begin{equation}
   \label{stftosph}
        \mathrm{Y}_{\ell m}=\mathcal{Y}_{\ell m}^L n_L~. 
   \end{equation}
    Where $\mathcal{Y}_{\ell m}^L$ are complex STF tensors. We may now note that \ref{philflatfar} can be re-expressed as
    \begin{equation}
    \label{philflatfar2}
        \begin{aligned}
            \phi_\ell=&\eta\sqrt{\eta} H^{(1)}_{\frac{3}{2}}(\Omega \eta) \sqrt{2 \pi \Omega}~  \Omega^{\ell}n_L (-1)^{\ell} \\
            &\times r^{-1 / 2}\left({}^{\mathrm{flat}}A_{\mathrm{reg}}^{L} J_{1/ 2+\ell}(\Omega r)+{}^{\mathrm{flat}}A_{\mathrm{irr}}^{L} Y_{1 / 2+\ell}(\Omega r)\right)
        \end{aligned}
    \end{equation}
    such that, 
    \begin{equation}
    \label{const2}
        {}^\mathrm{flat}A_{\mathrm{reg/irr}}^{L}=\sum\limits_{m=-\ell}^\ell\mathcal{Y}_{\ell m}^L{}^{\mathrm{flat}}A_{\mathrm{reg/irr}}^{\ell m}~.
    \end{equation}\\
    From the perspective of the worldline EFT for the SdS black hole, we should be identifying the coefficient ${}^\mathrm{flat}A^L_{\mathrm{reg}}/{}^\mathrm{flat}A^L_{\mathrm{irr}}$ with $C^L_{\mathrm{reg}}/ C^L_{\mathrm{irr}}$ where it should be understood that the ingoing boundary condition imposed at the horizon of SdS black hole will determine ${}^\mathrm{flat}A^L_{\mathrm{reg}}/{}^\mathrm{flat}A^L_{\mathrm{irr}}$. In what follows, we will explicitly illustrate how to compute these coefficients.\\
    
    Exploring the consequences of the near horizon physics on ${}^\mathrm{flat}A^L_{\mathrm{reg}}/{}^\mathrm{flat}A^L_{\mathrm{irr}}$ requires one to perform a near zone - far zone matching calculation \cite{Creci:2021rkz,Cardoso:2019nis,Cardoso:2004nk,Castro:2010fd}. But this task is not straightforward for the SdS black hole in the flat slicing coordinate (\ref{confbh}). Noting this, we will perform the near zone to far zone matching calculation in the static chart as is \ref{Matching the near zone with the far zone.} and identify the far zone solution. Following this, we will perform a chart transformation back to the flat slicing coordinate and compute the ratio ${}^\mathrm{flat}A^L_{\mathrm{reg}}/{}^\mathrm{flat}A^L_{\mathrm{irr}}$ as discussed in \ref{Going to flat slicing and the dS EFT.}.\\

    \subsection{Matching the near zone with the far zone.}
    \label{Matching the near zone with the far zone.}

    As discussed above, we will perform the near zone - far zone matching calculation  in the static chart  \cite{Creci:2021rkz,Cardoso:2019nis,Cardoso:2004nk,Castro:2010fd}, where the metric reads,
    \begin{equation}
            \begin{gathered}
                d s^2=-f(R)d T^2+f(R)^{-1} d R^2+R^2 d\Omega_2~,\\
                f(R)=1-\frac{2M}{R}-R^2 H^2~.
            \end{gathered}
    \end{equation}
    As the static patch of the SdS black hole is a subset of the region covered by the flat slicing coordinates, the static coordinates and the flat slicing coordinates can be seen to be related by \cite{Shiromizu:2001bg},
    \begin{equation}
    \label{stattoflatcor}
        \begin{gathered}
            R=a(t) r\left[1+\frac{M}{2a(t) r}\right]^{2 }, \\
            T=t+H \int^{R} \frac{R}{f(R)}\left(1-\frac{2 M}{R}\right)^{-1 / 2} d R,\\
            a(t)=e^{Ht},\quad -\frac{1}{H\eta}=e^{Ht}~.\\
        \end{gathered}
    \end{equation}
    Since the SdS spacetime in the static gauge has apparent killing symmetries associated with $T$ and $d\Omega_2$, we may expand the solution in Fourier and spherical harmonics and solve the radial part of the differential equation in two regions, the far and near zone, followed by a matching calculation in the intermediate region.

    \subsubsection{Far zone.}
    \label{Far Zone}

    In the far zone region, the spacetime of the SdS black hole should approach pure dS. For a SdS black hole, the far zone region can be observed to be characterized by $(M/R)<<R^2H^2$ and $R^2H^2\sim \mathcal{O}(1)$. In this region, the metric will look like pure de Sitter in the static chart and the radial part of the scalar field, ${}^\mathrm{far}\mathcal{R}(R)$ can be shown to obey the following differential equation.
    \begin{widetext}
    \begin{equation}
    \label{farzoneq}
            \begin{aligned}
                &R^2\left(1-H^2 R^2\right)^2 {}^\mathrm{far}\mathcal{R}^{\prime \prime}(R)+2R\left(1-H^2 R^2\right)\left(1-2 H^2 R^2\right) {}^\mathrm{far}\mathcal{R}^{\prime}(R)\\
                &\text{\hspace{1.5cm}}- \left(1-H^2 R^2\right)\left(\ell(\ell+1)-\frac{R^2 \omega^2}{1-H^2 R^2}\right) {}^\mathrm{far}\mathcal{R}(R)=0~.\\
            \end{aligned}
    \end{equation}
    \end{widetext}
    One may solve the above differential equation to obtain the far zone solution for the radial part of the SdS black hole ${}^\mathrm{far}\mathcal{R}(R)$ to be of the form
    \begin{widetext}
    \begin{equation}
    \label{solfarhor}
        \begin{aligned}
           {}^\mathrm{far}\mathcal{R}(R)=&~{}^\mathrm{stat}A^\ell_\mathrm{reg}R^{\ell}\left(1-H^2 R^2\right)^{\frac{-i \omega }{2 H}} \, _2F_1\left[\frac{1}{2} \left(\ell-\frac{i \omega }{H}\right),\frac{1}{2} \left(\ell-\frac{i \omega }{H}+3\right);\ell+\frac{3}{2};H^2 R^2\right]\\           
           &+{}^\mathrm{stat}A^\ell_\mathrm{irr}R^{-\ell-1}\left(1-H^2 R^2\right)^{\frac{-i \omega }{2 H}} {}_2F_1\left[-\frac{H(\ell+1)+i \omega }{2 H},1-\frac{\ell}{2}-\frac{i \omega }{2 H};\frac{1}{2}-\ell;H^2 R^2\right],
        \end{aligned}
    \end{equation}    
    \end{widetext}
    with ${}^\mathrm{stat}A^\ell_\mathrm{reg/irr}$ being constants that characterize the two linearly independent solutions in the far zone of SdS black hole.
    
    \subsubsection{Near zone.}
    \label{Near Zone}
    
    Here, we will solve the near zone scalar field equation and study the behaviour of the radial part of the scalar field in the near zone. To do this, we need to define the notion of the near zone carefully. Two possible notions of the near zone are available in the literature. The first notion is broadly based on the construction discussed in \cite{Creci:2021rkz,Cardoso:2019nis,Cardoso:2004nk,Castro:2010fd}. The second notion is discussed in earlier works as the near horizon region \cite{Castro:2010fd}. We shall solve the near zone radial differential equation within these two distinct notions of the near zone. These two notions are associated with two different regions of the spacetime. In this work, we have presented both calculations to explicitly illustrate certain subtleties present in Love number computations that may be related to the appropriate choice of the near zone.\\ 

    In what follows we shall consider the $\mathcal{O}(H^2M^2)$ corrections to the scalar field equation under the two different notions of the near zone.  

    \begin{center}
        \textbf{First notion.}
    \end{center}

    The first notion of near zone would be a minimal extension of the notion of the near zone region for Schwarzchild black holes. Here, due to the presence of a cosmological horizon and the associated length scale $1/H$, for a SdS black hole with black hole horizon radius $R_\mathrm{h}$; we impose $H(R-R_\mathrm{h})<<1$ along with $\omega (R-R_\mathrm{h})<<1$, \cite{Creci:2021rkz,Cardoso:2019nis,Cardoso:2004nk,Castro:2010fd} in the near zone region. This notion of the near zone should be understood as saying that the radial expanse of this region is much smaller than the length-scales, $1/H$ and $1/\omega$.\\
    
    Restricting the radial part of the field equation to this region allows us to replace $\omega R$ with $\omega R_\mathrm{h}$ and $HR$ with $HR_\mathrm{h}$. This will result in the following differential equation for the radial part of the perturbation.
    \begin{widetext}
        \begin{equation}
        \label{nearzoneq}
        \begin{aligned}
            &R^2\left(1-\frac{2M}{R}-H^2 R_\mathrm{h}^2\right)^2 {}^\mathrm{near}\mathcal{R}^{(1)\prime \prime}(R)+R\left(1-\frac{2M}{R}-H^2 R_\mathrm{h}^2\right)\left(2-\frac{2M}{R}-4 H^2 R_\mathrm{h}^2\right) {}^\mathrm{near}\mathcal{R}^{(1)\prime}(R)\\
            &\text{\hspace{1.5cm}}- \left(1-\frac{2M}{R}-H^2 R_\mathrm{h}^2\right)\left(\ell(\ell+1)-\frac{R_\mathrm{h}^2 \omega^2}{1-\frac{2M}{R}-H^2 R_\mathrm{h}^2}\right) {}^\mathrm{near}\mathcal{R}^{(1)}(R)=0~,\\
        \end{aligned}
    \end{equation}
    \end{widetext}
    ${}^\mathrm{near}\mathcal{R}^{(1)}(R)$ is the radial part of the scalar field in the first notion of the near zone. We will next attempt to solve the above differential equation to obtain ${}^\mathrm{near}\mathcal{R}^{(1)}(R)$ for a small SdS black hole.\\ 
    
    We shall now quantify the smallness of a SdS black hole through powers of $H^2M^2$; for a small SdS black hole, we can observe that \cite{Goswami:2022ylc},
    \begin{widetext}
        \begin{equation}
        \label{smallness}
        \begin{gathered}
            f(R)=1-\frac{2M}{R}-H^2R^2=\frac{H^2}{R}(R_\mathrm{c}-R)(R-R_\mathrm{h})(R+R_\mathrm{h}+R_\mathrm{c}),\\
            R_\mathrm{c}R_\mathrm{h}(R_\mathrm{c}+R_\mathrm{h})=\frac{2M}{H^2},\quad R_\mathrm{h}^2+R_\mathrm{c}^2+R_\mathrm{h}R_\mathrm{c}=\frac{1}{H^2},\quad 0\leq R_\mathrm{h}\leq R_\mathrm{c}\leq \frac{1}{H},\quad H^2M^2<<1,\\
            R_\mathrm{h}=M\left(2 + 8 H^2M^2+\mathcal{O}(H^3M^3)\right),\quad
            H R_\mathrm{c}=\big(1-HM-\frac{3}{2}H^2M^2+O(H^3M^3)\big)~.
        \end{gathered}
        \end{equation}
    \end{widetext}
    With the above notion of a small SdS black hole, we shall attempt to solve the scalar field equation \ref{nearzoneq} in the near zone region, perturbatively in $H^2M^2$.\\

    The perturbative solution for a small SdS black hole will be of the form,
    \begin{equation}
    \label{pertsol}
        \begin{aligned}
            {}^\mathrm{near}\mathcal{R}^{(1)}(R)=&{}^\mathrm{near}\mathcal{R}^{sh}(R)+4H^2M^2~h(R)\\
            &\text{\hspace{3.5cm}}+\mathcal{O}(H^3M^3)~.
        \end{aligned}
    \end{equation}
    Here ${}^\mathrm{near}\mathcal{R}^{sh}(R)$ is the leading order ingoing Schwarzchild solution \cite{Creci:2021rkz} and $h(R)$ is the correction to it at $\mathcal{O}(H^2M^2)$.\\
    
    After perturbatively expanding \ref{nearzoneq} in $H^2M^2$ and performing a variable redefinition of $\Tilde{f}=1-\frac{2M}{R}$, we can observe the leading order correction in the near zone, the function $h (f)$ to satisfy the following differential equation (dots representing derivatives with the variable $\Tilde{f}$).
    \begin{widetext}
    \begin{equation}
    \label{leading}
        \begin{gathered}
            \Tilde{f}(1-\Tilde{f})~\Ddot{h}+(1-\Tilde{f})~\dot{h}+\left(\frac{\omega^2(2M)^2(1-\Tilde{f})}{\Tilde{f}}-\frac{\ell(\ell+1)}{(1-\Tilde{f})}\right)~h=T(\Tilde{f}),\\
            T(\Tilde{f})=\frac{1}{\Tilde{f}(1-\Tilde{f})}\left(2\Tilde{f}(1-\Tilde{f})^2~{}^\mathrm{near}\Ddot{\mathcal{R}}^{sh}+(1-\Tilde{f}^2){}^\mathrm{near}\dot{\mathcal{R}}^{sh}-\left(8M^2\omega^2+\ell(\ell+1)\right){}^\mathrm{near}\mathcal{R}^{sh}\right),\\
           {}^\mathrm{near}\mathcal{R}^{sh}(\Tilde{f})= A \Tilde{f}^{2 i M\omega}(1-\Tilde{f})^{\ell+1}{}_2F_1[1+\ell+4iM\omega,\ell+1,1+4iM\omega,\Tilde{f}]~.
        \end{gathered}
    \end{equation}
    \end{widetext}
    We can now solve \ref{leading} using the method of variation of parameters. This allows us to obtain a formal solution for $h$ using the two linearly independent solutions to the homogenous part of \ref{leading} and the specific source term $T(\Tilde{f})$.\\ 
    
    For the dominant mode of the scalar perturbations having $\ell=0$, the formal solution for \ref{leading}, obtained using the method of variation of parameters $\left(h_{\ell=0}(\Tilde{f})\right)$ can be explicitly expressed as
    \begin{widetext}
        \begin{equation}
        \label{0cor}
        \begin{aligned}
            h_{\ell=0}(\Tilde{f})=&\Tilde{f}^{2 i M\omega}(1-\Tilde{f})\Bigg\{\frac{2 (2 M \omega +i) \, _2F_1[1,4 i M \omega -1;4 i M \omega ;\Tilde{f}]-4 M \omega -3 i}{2 (1-\Tilde{f}) \Tilde{f} (4 M \omega +i)}\\
            &\text{\hspace{6cm}}-\frac{2 M \omega  \, _2F_1[2,4 i M \omega -1;4 i M \omega ;\Tilde{f}]}{(1-\Tilde{f}) \Tilde{f} (4 M \omega +i)}\\
            &\text{\hspace{3cm}}-\frac{i \left(\frac{4 M \omega }{\Tilde{f}-1}+\frac{4 M \omega +i}{\Tilde{f}}+2 \log (1-\Tilde{f}) (2 M \omega -i)-2 \log (\Tilde{f}) (2 M \omega -i)\right)}{2 (1-\Tilde{f})}\Bigg\}~.
        \end{aligned}
    \end{equation}
    \end{widetext}
    Looking at the above equation, it might seem like it is divergent at $R=2M$, but this is not an issue; as discussed earlier, using \ref{smallness}, the small SdS black hole horizon $R_\mathrm{h}$ is shifted outside 2M.\\

    \begin{center}
        \textbf{Second notion.}
    \end{center}
        
    We shall construct the second notion of the near zone appropriately refered to as the near horizon region \cite{Castro:2010fd} by first writing the radial differential equation in the form,
    \begin{equation}
       \begin{gathered}
            \Delta\frac{d}{dR}\left[\Delta \frac{d~\mathcal{R}(R)}{dR}\right]+\left(R^4\omega^2-\ell(\ell+1)\right)\mathcal{R}(R)=0,\\
            \Delta=R^2f(R)~.
       \end{gathered}
    \end{equation}
    Now upon defining the variable $z=(R-R_\mathrm{h})/R_\mathrm{h}$, we have
    \begin{equation}
        \begin{aligned}
            \Delta
            &= H^2R_\mathrm{h}^2(R_\mathrm{c}-R_\mathrm{h})(R_\mathrm{c}+2R_\mathrm{h})z(1+z)\left(1-\frac{R_\mathrm{h}}{R_\mathrm{c}-R_\mathrm{h}}z\right)\\
            &\text{\hspace{2.5cm}}\times \left(1+\frac{R_\mathrm{h}}{R_\mathrm{c}+2R_\mathrm{h}}z\right)~.
        \end{aligned}
    \end{equation}
    We shall now note that we can get closer or further away from the horizon of the SdS black hole by dictating how small or large the variable $z$ is. Here, we shall define the second notion of the near zone as the region where we have $z$ to be small such that $\mathcal{O}(z^3)$ and higher powers of $z$ may be ignored in the expression for $\Delta$. Resulting in,
    \begin{equation}
        \begin{gathered}
        \Delta= H^2R_\mathrm{h}^2(R_\mathrm{c}-R_\mathrm{h})(R_\mathrm{c}+2R_\mathrm{h})z(1+\alpha z),\\
        \alpha=1-\frac{R_\mathrm{h}}{R_\mathrm{c}-R_\mathrm{h}}+\frac{R_\mathrm{h}}{R_\mathrm{c}+2R_\mathrm{h}}~.
        \end{gathered}
    \end{equation}
    In addition to this, similar to Schwarzchild black holes, we also demand $\omega(R-R_\mathrm{h})<<1$. So, we can write the radial differential equation in the near zone in terms of the variable $z$ as
    \begin{widetext}
    \begin{equation}
    \label{2ndnearzonebig}
            \begin{gathered}
            z(1+\alpha z)\frac{d}{dz}\left[z(1+\alpha z) \frac{d}{dz}\left[~  {}^{\text{near}}\mathcal{R}^{(2)}(z)\right]\right]
            +\left(\frac{\omega^2R_\mathrm{h}^2}{H^4(R_\mathrm{c}-R_\mathrm{h})^2(R_\mathrm{c}+2R_\mathrm{h})^2}
            -\frac{\ell(\ell+1) z(1+\alpha z)}{H^2(R_\mathrm{c}-R_\mathrm{h})(R_\mathrm{c}+2R_\mathrm{h})}\right) {}^{\text{near}}\mathcal{R}^{(2)}(z)=0.
            \end{gathered}
    \end{equation}
    \end{widetext}
    Where ${}^{\text{near}}\mathcal{R}^{(2)}$ is the radial part of the scalar field in the second notion of near zone. Next, we will restrict to a small SdS black hole, with smallness quantified through powers of $H^2M^2$. Upon keeping only terms upto $\mathcal{O}(H^2M^2)$ and $\mathcal{O}(M\omega)$, \ref{smallness} can be used to write \ref{2ndnearzonebig} in terms of the variable $y=\alpha z$ as
    \begin{widetext}
        \begin{equation}
            \begin{gathered}
            y(1+y)\frac{d}{dy}\left[y(1+y) \frac{d}{dy}\left[{}^{\text{near}}\mathcal{R}^{(2)}(y)\right]\right]
            +\Big(4\omega^2 M^2
            -\ell(\ell+1)(1+24H^2M^2) y(1+y)\Big) {}^{\text{near}}\mathcal{R}^{(2)}(y)=0.
            \end{gathered}
        \end{equation}
         \end{widetext}
    The ingoing solution to the above differential equation can be seen to have the form
    \begin{equation}
        \label{2ndNearzone}
        \begin{aligned}
                  {}^{\text{near}}R^{(2)}(y) =&(1+y)^{2iM\omega} y^{-2 i M\omega}\\
                  &{}_{2}F_{1}[-\Tilde{\ell},\Tilde{\ell}+1,1-4iM\omega,-y]
        \end{aligned}
    \end{equation}
    \begin{equation*}
                            \Tilde{\ell}=\ell+\frac{24\ell(\ell+1)}{2\ell+1}H^2M^2~.            
    \end{equation*}
    \subsubsection{Matching of near zone with far zone.}
    \label{matching}
    Having obtained the near zone and far zone solutions, our next task would be to perform a matching calculation of the near zone with the far zone \cite{Creci:2021rkz,Cardoso:2019nis,Cardoso:2004nk,Castro:2010fd}. We will perform this for the two notions of near zone mentioned above, for which the solutions were derived in \ref{Near Zone}.\\

    At this stage, we emphasise a key assumption regarding our calculation. In \ref{Near Zone}, we maintained the $\mathcal{O}(H^2M^2)$ terms for the near zone field equations. But in the far zone region of \ref{Far Zone}, we ignored all the $\mathcal{O}(H^2M^2)$ terms respecting the assumptions of the worldline EFT constructed \ref{deEFT}. However, we justify this analysis as the near zone region contains information regarding the behaviour of the compact object and thus Love numbers \cite{Hinderer:2007mb,Chia:2021,Cardoso:2017cfl}.\\
    
    First, we can note that the far zone region has an extra length scale $1/H$ in addition to $1/\omega$ so, to go to the matching region from the far zone, we need to take the limits $HR<<1$ and $\omega R<<1$ \cite{Creci:2021rkz,Cardoso:2019nis} of \ref{solfarhor}. This will result in
    \begin{equation}
    \label{fartonear2}
        \begin{aligned}
            {}^\mathrm{far}\mathcal{R} = \,&{}^\mathrm{stat}A^\ell_\mathrm{reg}\, R^{\ell}+{}^\mathrm{stat}A^\ell_\mathrm{irr}\, R^{-\ell-1}~.\\
        \end{aligned}
    \end{equation}
    Second, we can note that going to the matching region from the near zone would require going further away from the horizon of the black hole; thus, for a small SdS black hole, we should take the  $(M/R)<<1$ limit on the near-zone solution \cite{Creci:2021rkz,Cardoso:2019nis}.\\

    \begin{center}
        \textbf{Matching with the first notion.}
    \end{center}

    If we take the $(M/R)<<1$ limit on the solution obtained in the region specified by the first notion of the near zone as in \ref{pertsol}. We will get,    
    \begin{widetext}
        \begin{equation}
        \label{neartofar}
        \begin{aligned}
            {}^\mathrm{near}\mathcal{R}^{(1)}&=\left(\frac{\Gamma(-2 \ell-1) \Gamma\left(1+ 4iM\omega\right)}{\Gamma(-\ell) \Gamma\left(4 iM\omega-\ell\right)} +4H^2M^2~c^\mathrm{irr}_\ell\right)\left(\frac{2M}{R}\right)^{(\ell+1)} \\
            &\text{\hspace{1.25cm}}\quad+ \left(\frac{\Gamma(2 \ell+1) \Gamma\left(1+ 4iM\omega\right)}{\Gamma\left(\ell+1\right) \Gamma\left(\ell+1+ 4iM\omega\right)}+4H^2M^2~c^\mathrm{reg}_\ell\right)\left(\frac{R}{2M}\right)^{ \ell}.
        \end{aligned}
        \end{equation}
    \end{widetext}
    Where $c^\mathrm{reg}_\ell~\&~c^\mathrm{irr}_\ell$ are corrections that should arise at leading order in $H^2M^2$ for a small SdS black hole upon evaluating the solution for \ref{leading}. Comparing \ref{fartonear2} with \ref{neartofar} and identifying the powers of $R$ in the matching region, we may write 
    \begin{equation}
    \label{statcoefratio}
    \begin{gathered}
    \frac{{}^\mathrm{stat}A^\ell_\mathrm{irr}}{{}^\mathrm{stat}A^\ell_\mathrm{reg}}\Bigg|^{(1)}=(2M)^{2\ell+1}\frac{\left(\gamma_1 +4H^2M^2~c^\mathrm{irr}_\ell\right)}{\left(\gamma_2+4H^2M^2~c^\mathrm{reg}_\ell\right)},\\
    \gamma_1=\frac{\Gamma(-2 \ell-1) \Gamma\left(1+ 4iM\omega\right)}{\Gamma(-\ell) \Gamma\left(4 iM\omega-\ell\right)},\\
    \gamma_2=\frac{\Gamma(2 \ell+1) \Gamma\left(1+ 4iM\omega\right)}{\Gamma\left(\ell+1\right) \Gamma\left(\ell+1+ 4iM\omega\right)}.
    \end{gathered}
    \end{equation}
     For the dominant $\ell=0$ mode of a small SdS black hole, we are able to compute $c^\mathrm{reg}_0~\&~c^\mathrm{irr}_0$ using \ref{pertsol}, \ref{leading} and \ref{0cor} to be,\\
    \begin{equation}
    \label{0cor1}
        \begin{gathered}
            c^\mathrm{irr}_0=-\frac{1}{2},\quad c^\mathrm{reg}_0=-\frac{1}{2}+\frac{i}{4 M \omega }~.
        \end{gathered}
    \end{equation}
    Where $c^\mathrm{irr}_0$ and $c^\mathrm{reg}_0$ have been computed under the assumption of a small SdS black hole, that is, we can only keep terms proportional to $H^2M^2$ and $M\omega$ all higher order terms have been ignored while going to the matching region from the near zone.

    \begin{center}
        \textbf{Matching with the second notion.}
    \end{center}
    
    Now, instead of using \ref{pertsol} obtained from the first notion of the near zone. If we take the $M/R<<1$ limit on the near zone solution in the region specified by the second notion of the near zone, given by \ref{2ndNearzone}. We will get 
    \begin{widetext}
    \begin{equation}
    \label{nearfar2}
        \begin{aligned}
            {}^{\text{near}}\mathcal{R}^{(2)}=&\frac{\Gamma(1-2 i \omega)\Gamma(2\Tilde{\ell}+1)}{\Gamma(\Tilde{\ell}+1)\Gamma(1+\Tilde{\ell}-2 i\omega)}\left(1-H^2 M^2 \left(16 \ell-\frac{24 \ell (\ell+1) \log \left(\frac{R}{2 M}\right)}{2 \ell+1}\right)\right)\left(\frac{R}{2M}\right)^\ell\\
            &\hspace{3cm}+\frac{\Gamma(1-2i\omega)\Gamma(-1-2\Tilde{\ell})}{\Gamma(-\Tilde{\ell})\Gamma(-\Tilde{\ell}-2i\omega)}\left(1-H^2 M^2 \left(\frac{24 \ell(\ell+1) \log \left(\frac{R}{2 M}\right)}{2 \ell+1}-16 \ell-16\right)\right)\left(\frac{2M}{R}\right)^{\ell+1},
        \end{aligned}
    \end{equation}
    \end{widetext}
    where $\Tilde{\ell}$ is given by \ref{2ndNearzone}.\\
    
    Following this, we may identify the powers of $R$ in the matching region by comparing \ref{fartonear2} and \ref{nearfar2} to obtain an equation analogous to \ref{statcoefratio}, but based on the second notion of near zone. The ratio of the coefficients can be seen to be 
    \begin{widetext}
    \begin{equation}  
    \label{2ndnotionratio}
        \begin{gathered}       \frac{{}^\mathrm{stat}A^\ell_\mathrm{irr}}{{}^\mathrm{stat}A^\ell_\mathrm{reg}}\Bigg|^{(2)}=\frac{(2M)^{2\ell+1}\Gamma(\ell+1)\Gamma(-1-2\ell)\Gamma(\ell+1-4i\omega)}{\Gamma(2\ell+1)\Gamma(-\ell)\Gamma(-\ell+4iM\omega)}\left(1+H^2 M^2 \left(16+32 \ell+\frac{24 \ell (\ell+1) \Psi}{2 \ell+1}\right)\right),\\
        \Psi=\psi(-\ell-4 i M \omega )+\psi(\ell-4 i M \omega +1)+\psi(-\ell)+\psi(\ell+1)-2 \psi(2 \ell+1)-2 \psi(-2 \ell-1)-2 \log \left(\frac{R}{2 M}\right),\\
        \quad \psi(x)=\Gamma'(x)/\Gamma(x)~.
        \end{gathered}
    \end{equation}
    \end{widetext}
    An interesting observation regarding the above equation is that identifying the powers of $R$ when working with the second notion of near zone will result in ${}^\mathrm{stat}A^\ell_\mathrm{irr}/{}^\mathrm{stat}A^\ell_\mathrm{reg}~|^{(2)}$ having a $\log(R/2M)$ term in it. Some earlier works have also noted $\log$ terms, as in \ref{nearfar2} appearing in alternate theories of gravity \cite{DeLuca:2022tkm,Katagiri:2023umb}. Such $\log$ terms in the case of Schwarzchild black holes are interpreted as a consequence of classical RG flow \cite{Kol:2011vg,Ivanov:2022hlo,Hui:2020xxx}. \\
    
    One may criticise the identification of the powers of $R$ in \ref{fartonear2} in the matching region with the same in \ref{nearfar2} due to the $\log$ terms. However, we have performed this, as some earlier works suggest such terms to arise in the expression of Love numbers \cite{DeLuca:2022tkm,Perry:2023wmm}. In the next section, when we explicitly express the Love number, we will note that upon working with the second notion of near zone, we will necessarily have such $\log(R/2M)$ terms for SdS black hole Love numbers when $\ell\neq 0$.\\
    
    We would like to mention that the above analysis has been performed in the static chart of a SdS black hole. However, the dS worldline EFT and the definition of Love numbers employ the flat slicing coordinates as in \ref{Flat slicing coordinates for SdS and the EFT on the Poincaré patch of dS}. The next step is to go to the flat slicing coordinates in order to identify the scalar field in the asymptotic region of the metric \ref{confbh} with the EFT scalar field of \ref{deEFT} described in the Poincaré patch of pure de Sitter.

    \subsection{Going to flat slicing and the worldline EFT on dS.}
    \label{Going to flat slicing and the dS EFT.}
    
    After performing a matching calculation for the SdS black hole from the near zone to the far zone in the static chart, our next goal would be to go to the flat slicing coordinate of the SdS black holes, as discussed in \ref{Flat slicing coordinates for SdS and the EFT on the Poincaré patch of dS}, where we can perform the matching of the coefficients in the far zone of the black hole with the coefficients of the dS worldline EFT as in \ref{deEFT}.\\ 
    
    To go to the flat slicing coordinates from the static coordinates, we may employ \ref{stattoflatcor}, where $t$ is the cosmological time. Observe that the coordinate transformation in \ref{stattoflatcor} is greatly simplified when $(M/R)<<1$. So we first plug in the expression for the static coordinates from \ref{stattoflatcor} in the region where $(M/R)<<1$, into the \ref{solfarhor}. Then we find a region on the manifold in terms of the coordinates $(r,t)$ where the ratios ${}^{\mathrm{flat}}A^\ell_{\mathrm{irr}}/~{}^{\mathrm{flat}}A^\ell_{\mathrm{reg}}$ and ${}^\mathrm{stat}A^\ell_\mathrm{irr}/~{}^\mathrm{stat}A^\ell_\mathrm{reg}$ can be related by comparison with \ref{philflatfar}.\\ 
    
    Once we find a relation between the ratios in one region of the static patch, this relation should hold everywhere in the static patch, as these ratios are constants specifying the solution and should be the same throughout the static patch. Using the above-discussed procedure, coordinate invariance will result in the following relation between ${}^{\mathrm{flat}}A^\ell_{\mathrm{irr}}/~{}^{\mathrm{flat}}A^\ell_{\mathrm{reg}}$ and ${}^\mathrm{stat}A^\ell_\mathrm{irr}/~{}^\mathrm{stat}A^\ell_\mathrm{reg}$.
    \begin{equation}
        \frac{{}^{\mathrm{flat}}A^\ell_{\mathrm{irr}}}{{}^{\mathrm{flat}}A^\ell_{\mathrm{reg}}}=-\frac{\pi}{\Gamma(\frac{2\ell+3}{2})\Gamma(\frac{2\ell+1}{2})}\left(\frac{\Omega}{2}\right)^{2\ell+1}\frac{{}^\mathrm{stat}A^\ell_\mathrm{irr}}{{}^\mathrm{stat}A^\ell_\mathrm{reg}}.
    \end{equation}
    Observe that spherical symmetry implies ${}^\mathrm{flat}A^{\ell m}_\mathrm{reg/irr}$ of \ref{const1} will be independent of $m$ and \ref{const2} will result in,
    \begin{equation}
        \frac{{}^\mathrm{flat}A_\mathrm{irr}^{L}}{{}^\mathrm{flat}A_\mathrm{reg}^{L}}=\frac{{}^\mathrm{flat}A_\mathrm{irr}^{\ell}}{{}^{\mathrm{flat}}A_{\mathrm{reg}}^{\ell}}~.
    \end{equation}
    As discussed in \ref{Flat slicing coordinates for SdS and the EFT on the Poincaré patch of dS}, the matching of the microscopic description with the macroscopic dS worldline EFT, requires ${}^\mathrm{flat}A_\mathrm{irr}^{L}/~{}^\mathrm{flat}A_{\mathrm{reg}}^{L}=C^L_{\mathrm{irr}}/~C^L_{\mathrm{reg}}$. This means that for a comoving SdS black hole, the normalised tidal response of \ref{response}, ${}^{SdS}\Tilde{F}^\omega_\ell$ can be expressed in terms of  ${}^\mathrm{stat}A^\ell_\mathrm{irr}/~{}^\mathrm{stat}A^\ell_\mathrm{reg}$ to be
    \begin{equation}
    \label{SdSresponsereg}
        \begin{aligned}
            {}^{SdS}\Tilde{F}^\omega_\ell&=-\frac{4 \pi^{1 / 2}}{2^{\ell}}\left(\frac{2}{\Omega}\right)^{1+2 \ell} \Gamma\left(\frac{2\ell+3}{2}\right)\frac{C^L_{\mathrm{irr}}}{C^L_{\mathrm{reg}}}\times e^{H t(2\ell+1)}\\
            &=\frac{-\pi^{\frac{3}{2}}}{2^{\ell-2}~\Gamma(\frac{2\ell+1}{2})}\frac{{}^\mathrm{stat}A^\ell_\mathrm{irr}}{{}^\mathrm{stat}A^\ell_\mathrm{reg}}\times e^{H t(2\ell+1)}~.
        \end{aligned}
    \end{equation}
    Even though, in the context of this paper, \ref{SdSresponsereg} is for a SdS black hole. The arguments above are valid for all spherically symmetric comoving compact objects on a de Sitter background whose exterior spacetime is described by \ref{confbh}. As a result the above expression can be used for any such spherically symmetric compact object, where we will have to compute ${}^\mathrm{stat}A^\ell_\mathrm{irr}/{}^\mathrm{stat}A^\ell_\mathrm{reg}$ separately for each such compact object depending on the appropriate boundary conditions on their surface \cite{Cardoso:2017cfl,Hinderer:2007mb}.\\
    
    We will now note that the near zone - far zone matching calculation in \ref{Matching the near zone with the far zone.} has resulted in two different expressions for ${}^\mathrm{stat}A^\ell_\mathrm{irr}/~{}^\mathrm{stat}A^\ell_\mathrm{reg}$ which are ${}^\mathrm{stat}A^\ell_\mathrm{irr}/~{}^\mathrm{stat}A^\ell_\mathrm{reg}~|^{(1)}$ and ${}^\mathrm{stat}A^\ell_\mathrm{irr}/~{}^\mathrm{stat}A^\ell_\mathrm{reg}~|^{(2)}$ depending on the use of the first or the second notion of the near zone respectively. Next, we will explicitly write down the expression for the scalar Love numbers for SdS black holes as computed using these two distinct notions of the near zone.

    \subsection{SdS Love numbers}
    \label{SdS Love numbers}
    Here, we shall explicitly write down the expression for SdS Love numbers. We will first use the definition of Love numbers developed using the dS worldline EFT and \ref{SdSresponsereg}; this is the Love number of the SdS black hole as measured by a distant comoving observer. Following this, we will also comment on a functional notion of Love numbers in the static coordinates of a SdS black hole.
    
    \subsubsection{Love numbers for a comoving observer.}
    \label{Love numbers for a comoving observer.}
    
    Using the microscopic computation of a scalar field in a SdS background as discussed above, we are able to identify the Love numbers of a SdS black hole as observed by a comoving observer through \ref{SdSresponsereg} using a worldline EFT framework. Further in \ref{deEFT} within the framework of a worldline EFT, we had argued that the expression \ref{response} can be understood as a well-defined notion of tidal Love numbers for comoving compact objects in terms of scattering coefficients as measured by a distant comoving observer on a de Sitter background. Here, we shall explicitly express the SdS Love number up to $\mathcal{O}(H^2M^2)$.\\
    
    To get the explicit form of the Love number, we shall plug in the value for ${}^\mathrm{stat}A^\ell_\mathrm{irr}/~{}^\mathrm{stat}A^\ell_\mathrm{reg}$ into \ref{SdSresponsereg}. However, since the use of two different notions of the near zone has resulted in two different expressions for ${}^\mathrm{stat}A^\ell_\mathrm{irr}/~{}^\mathrm{stat}A^\ell_\mathrm{reg}$ which are ${}^\mathrm{stat}A^\ell_\mathrm{irr}/~{}^\mathrm{stat}A^\ell_\mathrm{reg}~|^{(1)}$ and ${}^\mathrm{stat}A^\ell_\mathrm{irr}/~{}^\mathrm{stat}A^\ell_\mathrm{reg}~|^{(2)}$ respectively. The associated response coefficients ${}^{SdS}\Tilde{F}^{\omega(1)}_\ell$ and ${}^{SdS}\Tilde{F}^{\omega(2)}_\ell$ are also different. Observe that
    \begin{equation}
    \label{SdSresponsereg1}
        \begin{aligned}{}^{SdS}\Tilde{F}^{\omega(1)}_\ell=\frac{-\pi^{\frac{3}{2}}}{2^{\ell-2}~\Gamma(\frac{2\ell+1}{2})}\frac{{}^\mathrm{stat}A^\ell_\mathrm{irr}}{{}^\mathrm{stat}A^\ell_\mathrm{reg}}\Bigg|^{(1)}\times e^{H t(2\ell+1)}~.
        \end{aligned}
    \end{equation}
    Now, expanding ${}^\mathrm{stat}A^\ell_\mathrm{irr}/~{}^\mathrm{stat}A^\ell_\mathrm{reg}~|^{(1)}$ from \ref{statcoefratio} and keeping terms up to $\mathcal{O}(H^2M^2)$ for a small SdS black hole will result in
    \begin{equation}
    \label{responseregSdS1}
        \begin{aligned}       {}^{SdS}\Tilde{F}^{\omega (1)}_\ell=\left({}^\mathrm{sch}\Tilde{F}^\omega_\ell+4H^2M^2L^{\omega (1)}_\ell\right)\times e^{H t(2\ell+1)}~. 
        \end{aligned}
    \end{equation}
    Where ${}^\mathrm{sch}\Tilde{F}^{\omega}_\ell$ is the normalised response coefficient of a Schwarzschild black hole as defined in \ref{response} and $L^{\omega (1)}_\ell$ is the expected correction, which arises at $\mathcal{O}(H^2M^2)$. For the dominant mode of the scalar perturbations ($\ell=0$), we can explicitly evaluate $L^{\omega (1)}_0$ using \ref{statcoefratio} and \ref{0cor1} to be
    \begin{equation}
        L^{\omega (1)}_0=8M\pi~.
    \end{equation}
    This means that for a small SdS black hole, the normalised Love number, as observed by a comoving observer for the dominant mode of a scalar perturbation under the first notion of near zone, looks like,
    \begin{equation}
    \label{explicitLove}
    \begin{aligned}
        \Tilde{k}^{\omega(1)}_0&=\mathrm{Re}\left[{\Tilde{F}^{\omega (1)}_0}\right]\\
        &=32\pi H^2M^3e^{Ht}.
    \end{aligned}
    \end{equation}
    Similarly, if we adopt the second notion of near zone from \ref{SdSresponsereg} we have 
    \begin{equation}
    \label{2ndSdSresponsereg}
        \begin{aligned}
        {}^{SdS}\Tilde{F}^{\omega(2)}_\ell&=\frac{-\pi^{\frac{3}{2}}}{2^{\ell-2}~\Gamma(\frac{2\ell+1}{2})}\frac{{}^\mathrm{stat}A^\ell_\mathrm{irr}}{{}^\mathrm{stat}A^\ell_\mathrm{reg}}\Bigg|^{(2)}\times e^{H t(2\ell+1)}.\\
        \end{aligned}
    \end{equation}
    If we go ahead and plug in the expression for ${}^\mathrm{stat}A^\ell_\mathrm{irr}/~{}^\mathrm{stat}A^\ell_\mathrm{reg}~|^{(2)}$ within the second notion of near zone from \ref{2ndnotionratio} we will get 
    \begin{equation}
    \begin{aligned}
            {}^{SdS}\Tilde{F}^{\omega(2)}_\ell&={}^\mathrm{sch}\Tilde{F}^\omega_\ell\left[1+H^2 M^2 \left(16+32 \ell+\frac{24 \ell (\ell+1) \Psi}{2 \ell+1}\right)\right]\\
            &\hspace{4cm}\times e^{H t(2\ell+1)}~.
    \end{aligned}
    \end{equation}
    Further, the normalised Love number will be 
    \begin{equation}
    \label{explicitLove2}
    \begin{aligned}                 \Tilde{k}^{\omega(2)}_\ell&=\mathrm{Re}\left[{\Tilde{F}^{\omega (2)}_0}\right]\\
        &=-\frac{24 \ell (\ell+1) }{2 \ell+1}~\mathrm{Im}\left[{}^\mathrm{sch}\Tilde{F}^\omega_\ell\right]~\mathrm{Im}\left[\Psi\right] e^{H t(2\ell+1)}.
    \end{aligned}
    \end{equation}
    Where $\Psi$ has the form given in \ref{2ndnotionratio} and is not always real; however, it should be noted that the two expressions for the SdS Love number cannot be simultaneously correct; only one of the two is correct. The reason for getting two different expressions is the use of two different notions of near zone.\\

    Even though the first notion of near zone is the simplest and most straightforward definition of a near zone, as it is a straightforward extension of the one used in the Schwarzchild case \cite{Creci:2021rkz,Cardoso:2019nis,Cardoso:2004nk,Castro:2010fd} , but with an additional length scale $1/H$. However, we do not find any reason to completely discard the second notion either, as it also quantifies a region of the background metric at a certain degree of proximity to the black hole horizon. The disagreement between the two different computations may indicate certain intricacies regarding the computation of Love numbers for compact objects. However, both computations, either with the first or second notion of near zone, indicate a non-zero value for the Scalar SdS Love numbers.
    
    \subsubsection{Love numbers in static coordinates.}
    Looking at the far zone limit of the near zone solution in the static chart from \ref{Matching the near zone with the far zone.} it might seem that taking the $M/R<<1$ limit on the near zone solution and taking the ratios of the coefficients of the growing and the decaying terms is sufficient to define the response coefficient of the compact object based on an analogy with the asymptotically flat case \cite{Zhang:1986cpa,Hinderer:2007mb,Binnington:2009bb}; however one should tread with caution when it comes to this definition. This is because the definition of Love numbers in the asymptotically flat case is motivated through the analogy with the Newtonian notion, where the response of the compact object is quantified by the coefficient of the $R^{-\ell-1}$ while the presence of the source is signified through the coefficient of $R^\ell$. Such an analogy may not be possible if the spacetime is not asymptotically flat.\\ 
    
    However, looking at \ref{SdSresponsereg} it is clear that the Love number for a spherically symmetric compact object as observed by a distant comoving observer(in terms of ingoing and outgoing scattering coefficients) is specified entirely through the ratio ${}^\mathrm{stat}A^\ell_\mathrm{irr}/{}^\mathrm{stat}A^\ell_\mathrm{reg}$. This suggests that one may very well use ${}^\mathrm{stat}A^\ell_\mathrm{irr}/{}^\mathrm{stat}A^\ell_\mathrm{reg}$ as a functional definition of the response coefficient for asymptotically de Sitter spacetimes in the static coordinates. So we define a functional notion of response coefficient for a spherically symmetric comoving compact object, $\Tilde{F}^{\omega (s)}_\ell$  as the ratio\\
    \begin{equation}
    \label{StaticLovedef}
         \Tilde{F}^{\omega (s)}_\ell=\frac{{}^\mathrm{stat}A^\ell_\mathrm{irr}}{{}^\mathrm{stat}A^\ell_\mathrm{reg}}~.
    \end{equation}
    Using the above notion of response coefficients for a comoving compact object, in terms of the static coordinates on a SdS background, one can use \ref{statcoefratio} or \ref{2ndnotionratio} depending on the first or the second notion of near zone to observe a nonzero Love number for the SdS black holes.
    
    \section{Summary and Discussion}
    \label{Summary and Discussion}

    The tidal response of a body against an external tidal field $\mathcal{E}_L(x^0)$, is quantified using the tidal response function $F_\ell(x^0)$, which is defined through the relationship
    \begin{equation}
        Q_L(x^0)=-F_\ell(x^0)\mathcal{E}_L(x^0)~,
    \end{equation}
    where $Q_L(x^0)$  is the induced multipole moment of the body. The real part of the tidal response function constitutes the Love number of the body, $k_\ell(x^0)=\mathrm{Re}\left[F_\ell(x^0)\right]$. In general, we expect the tidal response and the multipole moment to be functions of the coordinate time($x^0$) if the vector field associated with the time is not Killing. This is relevant to our construction, where we consider asymptotically deSiter compact objects.\\
    
    Since the idea of tidal Love numbers is not straightforward if the spacetime is not asymptotically flat, we constructed a worldline Effective Field Theory (EFT) for compact objects on a non-flat background in order to define Love numbers for asymptotically non-flat scenarios. Within the framework of the worldline EFT, we define the black hole Love numbers in terms of the scattering coefficients associated with the perturbing fields, as measured by the distant observer.\\
    
    Subsequently, we focused on compact objects on a de Sitter background where the body interacts with a background scalar field by coupling with its multipole moments as described in \ref{totalact} and \ref{tidact}. Working in the Poincaré patch for the dS spacetime, we express its response function as
    \begin{equation}
     {F}_\ell(t)={K_\phi}\Tilde{F}_\ell(t)~.
    \end{equation}
    where $K_\phi$ is the scalar field coupling constant and $t$ is the cosmological time. $\Tilde{F}_\ell(t)$ is the normalised response function, and it may be expressed as
    \begin{equation}
    \label{res2}
        \Tilde{F}_\ell(t)=-\frac{4 \pi^{1 / 2}}{2^{\ell}}\left(\frac{2}{\Omega}\right)^{1+2 \ell} \Gamma\left(\frac{2\ell+3}{2}\right)\frac{C^L_{\mathrm{irr}}}{C^L_{\mathrm{reg}}}\times e^{H t(2\ell+1)}.
    \end{equation}
    Where we have, $H=\sqrt{\Lambda/3}$; $C^L_{\mathrm{irr}}~\&~C^L_{\mathrm{reg}}$ are constants characterising the compact object. From the perspective of the worldline EFT, these are constants which can be associated with the amplitudes of the ingoing and outgoing modes $(C^L_\mathrm{in},C^L_\mathrm{out})$ of the perturbation as observed by a distant comoving observer. The relation between $(C^L_{\mathrm{irr}}, C^L_{\mathrm{reg}})$ and $(C^L_\mathrm{in}, C^L_\mathrm{out})$ can be obtained by using the relation between Bessel functions and Hankel functions to be \ref{hanktobas}.\\

    While defining Love numbers for asymptotically dS spacetimes from a worldline EFT, we also noted certain features of a possibly non-dS background spacetime for which a similar approach could be used to define Love numbers. We also comment that the definition of Love numbers within the framework of a worldline EFT in terms of the scattering coefficients is observer-dependent as the scattering coefficients are themselves dependent on the observer. This is, however, not true if we are restricting to distant free-falling observers in the context of asymptotically flat spacetimes.\\

    In asymptotically flat spacetimes, we may use Newtonian analogy and define Love numbers using the asymptotic fall of the perturbations. But, for asymptotically non-flat cases, this notion is ambiguous, and we need to use the worldline EFT to define Love numbers. The worldline EFT setup for compact objects on a de Sitter background allows for a well-defined notion of Love numbers for asymptotically de Sitter compact objects in terms of the scattering coefficients of the perturbation, as observed by a distant observer. This can be thought of as an extension of the notion of Love numbers developed for asymptotically flat spacetimes within worldline EFT \cite{Creci:2021rkz,Bautista:2021wfy,Bernuzzi:2008rq}.\\

    From \ref{res2}, we can note that the response function is not independent of coordinate time, unlike in flat space. One can attribute this to the fact that this Love number expression is valid for a comoving observer whose time is not Killing.\\ 

    After developing a worldline EFT for comoving compact objects on the Poincaré patch of the de Sitter spacetime and using it to express the scalar Love number in terms of the ratio $C^L_{\mathrm{irr}}/~C^L_{\mathrm{reg}}$, we focused on a small SdS black hole of mass $M$, where an expansion in $H^2M^2$ quantifies the smallness. We computed the Love numbers of a small SdS black hole as measured by a distant comoving observer.\\
    
    The computation of SdS black holes involved a near zone - far zone matching calculation in the static chart of the SdS black hole. Following this, we went from the static chart to the flat slicing coordinates, where the metric reads \ref{confbh}. This allows for a matching with the worldline EFT, where we identified the ratio $C^L_{\mathrm{irr}}/~C^L_{\mathrm{reg}}={}^\mathrm{flat}A^L_{\mathrm{reg}}/{}^\mathrm{flat}A^L_{\mathrm{irr}}$, ${}^\mathrm{flat}A^L_{\mathrm{reg/irr}}$ being constants characterising the scalar field in the flat slicing coordinates.\\ 

    While computing the SdS Love numbers we used two different notions of the near zone, the first being the natural extension of the near zone for Schwarzchild black holes \cite{Creci:2021rkz,Cardoso:2019nis,Cardoso:2004nk,Castro:2010fd} with the additional cosmological length scale of $1/H$; the second being a notion of near zone quantified though how far we go from the horizon in powers of $(R-R_\mathrm{h})/R_\mathrm{h}$, which is actually a notion of the near-horizon region \cite{Castro:2010fd}. We have used the two different notions of the near zone to quantify proximity to the horizon to illustrate the intricacies in the Love number computation associated with the correct choice of the near zone.\\
    
    For a small SdS black hole, the normalised response function, under the first notion of the near zone, can be expressed as ${}^{SdS}\Tilde{F}^{\omega(1)}_\ell$ which reads.
    \begin{equation}
    \label{responseregSdS1}
        \begin{aligned}                             {}^{SdS}\Tilde{F}^{\omega(1)}_\ell=\left({}^\mathrm{sch}\Tilde{F}^\omega_\ell+4H^2M^2L^{\omega (1)}_\ell\right)\times e^{H t(2\ell+1)}~. 
        \end{aligned}
    \end{equation}
    With ${}^\mathrm{sch}\Tilde{F}^\omega_\ell$, being the normalised response function for Schwarzchild black hole. We have explicitly evaluated the expected correction $L^{\omega(1)}_\ell$ for the dominant mode of the scalar perturbation $(\ell=0)$ and observed that the response function is,
    \begin{equation}
    \label{responseregSdS1}
        \begin{aligned}
            {}^{SdS}\Tilde{F}^{\omega(1)}_0=\left\{{}^\mathrm{sch}\Tilde{F}^\omega_0+32\pi H^2M^3\right\}\times e^{H t}~. 
        \end{aligned}
    \end{equation}
    Resulting in the leading order normalised Love number under the first notion of the near zone, $\Tilde{k}^{\omega(1)}_0=\mathrm{Re}\left[{\Tilde{F}^{\omega(1)}_0}\right]$ to be
    \begin{equation}
       \Tilde{k}^{\omega(1)}_0(t)=32\pi H^2M^3\times e^{H t}~.
    \end{equation}
    Which is nonzero, unlike the asymptotically flat black holes of Einstein gravity \cite{Chia:2021,Ivanov:2022qqt,Creci:2021rkz,Kol:2011vg,Ivanov:2022hlo,Addressingissues:2023}.\\

    Instead, if we use the second notion of the near zone, we get
    \begin{equation}
    \begin{aligned}
            {}^{SdS}\Tilde{F}^{\omega(2)}_\ell&={}^\mathrm{sch}\Tilde{F}^\omega_\ell\left[1+H^2 M^2 \left(16+32 \ell+\frac{24 \ell (\ell+1) \Psi}{2 \ell+1}\right)\right]\\
            &\hspace{4cm}\times e^{H t(2\ell+1)}~,
    \end{aligned}
    \end{equation}
    where $\Psi$ is given by \ref{2ndnotionratio}, which clearly contains a $\log\left(R/2M\right)$ term. Such $\log$ terms were reported in earlier works when alternate theories of gravity were considered \cite{DeLuca:2022tkm,Katagiri:2023umb} and in the case of Schwarzchild black holes are interpreted as a consequence of classical RG flow \cite{Kol:2011vg,Ivanov:2022hlo,Hui:2020xxx}.\\

    Further, we can express the SdS normalised Love number when employing the second notion of the near zone, $\Tilde{k}^{\omega(2)}_0=\mathrm{Re}\left[{\Tilde{F}^{\omega(2)}_0}\right]$ to be
    \begin{equation}
     \Tilde{k}^{\omega(2)}_\ell=-\frac{24 \ell (\ell+1) }{2 \ell+1}~H^2M^2~\mathrm{Im}\left[{}^\mathrm{sch}\Tilde{F}^\omega_\ell\right]~\mathrm{Im}\left[\Psi\right] e^{H t(2\ell+1)}.
    \end{equation}
    One can clearly see that the Love number derived within the second notion of the near zone is distinct from the one derived from the first notion; this may be associated with an appropriate choice of the near zone being essential for computing the Love number of a compact object.\\

    We would also like to point out that a functional notion of Love numbers in the static coordinates for an asymptotically de Sitter compact object is, 
    \begin{equation}
    \label{StaticLovedef2}
         \Tilde{F}^{\omega (s)}_\ell=\frac{{}^\mathrm{stat}A^\ell_\mathrm{irr}}{{}^\mathrm{stat}A^\ell_\mathrm{reg}}~,
    \end{equation}
    ${}^\mathrm{stat}A^\ell_\mathrm{reg/irr}$ being coefficients characterising the scalar field in the static chart. We legitimise the validity of \ref{StaticLovedef2} as a measure of the tidal response of the compact object, as this ratio completely specifies the response coefficient measured by the comoving observer within the worldline EFT framework. These ratios were computed for the SdS black hole with the first and the second notion of the near zone and are given by in \ref{statcoefratio} and \ref{2ndnotionratio}, respectively.\\
    
    An interesting observation regarding the computation of the black hole Love number presented here is the use of an ingoing boundary condition at the black hole horizon; imposing an ingoing condition necessarily requires the perturbation frequency to be non-zero. However, one can go to the static limit of the Love number by taking the $ \omega\longrightarrow 0 $ limit. One may also obtain the static Love numbers by initially setting $ \omega=0 $ and solving the differential equation. Earlier works have demonstrated that these two types of calculations may result in different results, owing to the distinct branches of solution for the hypergeometric differential equation \cite{Nair:2022xfm,Chakraborty:2023zed}.\\
    
    Despite the calculations given in this work being for four dimensions, the calculations may be extended to account for higher dimensions using the machinery developed in \cite{Creci:2021rkz}; replicating the calculations in \ref{deEFT} on the Poincaré patch of the higher dimensional dS spacetime; and working with a higher dimensional SdS black hole instead to the four-dimensional one used in \ref{Love numbers of Schwarzschild de Sitter}. 
        
    
    \section{Conclusion}
    \label{Conclusion}
    
    In this paper, we used a worldline effective field theory framework for asymptotically de Sitter compact objects to define scalar Love numbers for comoving compact objects on a de Sitter background. The Love numbers can be defined using the scattering coefficients of the scalar field as observed by a far away comoving observer. As the comoving time is not Killing, we obtain a time-dependent expression for the Love numbers of these compact objects as measured by a comoving observer. Along the way, we also note the possibility of defining the Love number in a certain category of spacetimes in terms of scattering coefficients.\\

    We computed the $\mathcal{O}(\Lambda M^2)$ Love number for scalar perturbations of Schwarzchild de Sitter black hole having mass $M$ and cosmological constant $\Lambda$. In computing the Love numbers, we worked with two notions of the near zone; the first notion is an extension of the near zone notion from Schwarzchild black holes with an additional length scale introduced by the cosmological constant; the second notion of near zone is based on a quantification of the radial proximity to the black hole horizon $(R_\mathrm{h})$ in powers of $(R-R_\mathrm{h})/R_\mathrm{h}$. We note that the expressions for the Love number depend on the notion of near zone used, highlighting an ambiguity regarding the correct notion of near zone. However, both of the notions of near zone resulted in a nonzero value for the SdS Love numbers at $\mathcal{O}(\Lambda M^2)$.\\

    The nonzero value of the Schwarzchild de Sitter black hole scalar Love number shows that, even within the framework of Einstein gravity, objects with a horizon can have nonzero Love numbers. This has significant observational consequences as Love numbers are often considered to be a probe for the existence of horizons \cite{Cardoso:2017cfl,Nair:2022xfm,Katagiri:2023yzm,Katagiri:2023umb,DeLuca:2022tkm,Chakraborty:2023zed}.\\

    However it should be noted that $\Lambda M^2$ is negligible for astrophysical black holes, and looking for an observational consequence of an interaction of these two disparate length scales is incomplete without accounting for the matter and other effects on the measured Love number \cite{Gaur:2023hmk}. However, we argue that the calculations presented here may be more significant than the Love number computation for asymptotically flat black holes from an observational perspective.\\

    It would be interesting to extend our formalism to account for black hole spin and to other non-flat backgrounds, particularly asymptotically Anti-de Sitter (AdS) compact objects, and understand the tidal response of Schwarzschild/Kerr-AdS black holes. Further, it would be of interest to understand the tidal response of asymptotically non-flat black holes to metric perturbations.
    

    \section*{Acknowledgements}

    The authors are thankful to Rajes Ghosh for extensive discussions. We also thank Tanja Hinderer and Jan Steinhoff for comments on the draft. S.N. and S.S. also thank IACS, Kolkata, for hospitality, where part of this work was carried out. Research of S.C. is funded by the INSPIRE Faculty fellowship from DST, Government of India (Reg. No. DST/INSPIRE/04/2018/000893) and by the Start-Up Research Grant from SERB, DST, Government of India (Reg. No. SRG/2020/000409). The research of S. S. is supported by the Department of Science and Technology, Government of India, under the SERB CRG Grant (No. CRG/2020/004562). Research of S.N is supported by the Prime Minister's Research Fellowship (ID-1701653), Government of India. 


\bibliographystyle{plainnat}
\bibliographystyle{./utphys1}
\end{document}